\documentclass[twocolumn,twocolappendix]{aastex631}
\usepackage{amsmath}
\usepackage{amssymb}
\usepackage[utf8]{inputenc}
\usepackage[english]{babel}
\usepackage{amsfonts}

\newcommand{\p}{\partial}
\usepackage{graphics}
\usepackage{graphicx}
\usepackage{footnote}
\usepackage{txfonts}
\usepackage{lipsum}
\usepackage{float}
\usepackage{gensymb}
\usepackage{bm}

\DeclareSymbolFont{cmletters}{OML}{cmm}{m}{it}
\DeclareMathSymbol{v}{\mathalpha}{cmletters}{"76}

\definecolor{darkblue}{rgb}{0.0,0.0,0.3}
\hypersetup{colorlinks,breaklinks,
            linkcolor=darkblue,urlcolor=darkblue,
            anchorcolor=darkblue,citecolor=darkblue}

\newcommand{\rg}{r_\mathrm{g}}

\begin{document}

\title{Dual Jet Interaction, Magnetically Arrested Flows, and Flares\\ in Accreting Binary Black Holes}

\author[0000-0003-0220-5723]{Sean M. Ressler}
\affiliation{Canadian Institute for Theoretical Astrophysics, University of Toronto, Toronto, On, Canada M5S 3H8}

\author[0000-0002-5427-1207]{Luciano Combi}
\affiliation{Perimeter Institute for Theoretical Physics, Waterloo, Ontario N2L 2Y5, Canada}
\affiliation{Department of Physics, University of Guelph, Guelph, Ontario N1G 2W1, Canada}

\author[0000-0002-7301-3908]{Bart Ripperda}
\affiliation{Canadian Institute for Theoretical Astrophysics, University of Toronto, Toronto, On, Canada M5S 3H8}
\affiliation{David A. Dunlap Department of Astronomy, University of Toronto, 50 St. George Street, Toronto, ON M5S 3H4}
\affiliation{Department of Physics, University of Toronto, 60 St. George Street, Toronto, ON M5S 1A7}
\affiliation{Perimeter Institute for Theoretical Physics, Waterloo, Ontario N2L 2Y5, Canada}

\author[0000-0002-0491-1210]{Elias R. Most}
\affiliation{TAPIR, Mailcode 350-17, California Institute of Technology, Pasadena, CA 91125, USA}
\affiliation{Walter Burke Institute for Theoretical Physics, California Institute of Technology, Pasadena, CA 91125, USA}

\begin{abstract}
Supermassive binary black holes in galactic centers are potential multimessenger sources in gravitational waves and electromagnetic radiation. To find such objects, isolating unique electromagnetic signatures of their accretion flow is key.
With the aid of three-dimensional general-relativistic magnetohydrodynamic (GRMHD) simulations that utilize an approximate, semi-analytic, super-imposed spacetime metric, we identify two such signatures for merging binaries.
Both involve magnetic reconnection and are analogous to plasma processes observed in the solar corona.
The first, like colliding flux tubes that can cause solar flares, involves colliding jets that form an extended reconnection layer, dissipating magnetic energy and causing the two jets to merge. 
The second, akin to coronal mass ejection events, involves the accretion of magnetic field lines onto both black holes; these magnetic fields then twist, inflate, and form a trailing current sheet, ultimately reconnecting and driving a hot outflow.
We provide estimates for the associated electromagnetic emission for both processes, showing that they likely accelerate electrons to high energies and are promising candidates for continuous, stochastic, and/or quasi-periodic higher energy electromagnetic emission.
We also show that the accretion flows around each black hole can display features associated with the magnetically arrested state.
However, simulations with black hole spins misaligned with the orbital plane and simulations with larger Bondi radii saturate at lower values of horizon-penetrating magnetic flux than standard magnetically arrested disks, leading to weaker, intermittent jets due to feedback from the weak jets or equatorial flux tubes ejected by reconnecting field lines near the horizon. 
\end{abstract}

\keywords{Active galactic nuclei, Supermassive black holes, Accretion, Magnetohydrodynamical simulations, General relativity}

\section{Introduction}

Supermassive black hole (SMBH) binaries are expected to form as a natural consequence of galaxy mergers \citep{Kormendy2013}. 
The recent detection of an isotropic, low-frequency background gravitational wave signal by pulsar timing arrays (PTAs) provides strong observational evidence that such systems are relatively common \citep{Agazie2023,EPTA2023,Reardon2023}.
These observations, however, cannot yet resolve
individual in-spiralling binaries \citep{NANOGrav:2023pdq}. Future space-based gravitational wave missions like the Laser Interferometer Space Antenna (LISA, \citealt{Flanagan1998,Amaro2007,Berry2013}) promise to provide more precise localization for SMBH binary sources near and during merger.
Electromagnetically, while there are a handful of promising candidates \citep{Valtonen2008,Dotti2009,Liu2019,Jiang2022,Pasham2024,Kiehlmann:2024fji}, no observed galaxy has been fully confirmed to host a SMBH binary with close enough separation to be in the gravitational-wave regime.

In order to make progress, predictive modelling of electromagnetic emission provided by the accretion flows surrounding SMBH binaries is critical.
In particular, it is imperative to isolate emission mechanisms that are unique to binary accretion systems in order to distinguish them from active galactic nuclei (AGN) containing only a single SMBH, which are typically highly variable and red-noise dominated. 
One possibility is that the time-dependent gravitational potential of the binary can induce hydrodynamic periodicity in the flow, which can then correlate to periodicity in the otherwise stochastic thermal emission  (see, e.g., \citealt{DOrazio:2023rvl} for a recent review).
Another possibility is that there might be distinct non-thermal emission channels in SMBH binary accretion flows. 
The latter possibility will be the main focus of this work.

One recently proposed mechanism for unique non-thermal emission in SMBH binary accretion flows is magnetic reconnection between the interacting jets powered by the accretion of magnetic fields onto two rapidly spinning black holes (\citealt{Gutierrez2024}, see also \citealt{Palenzuela:2010nf}).
Magnetic reconnection, where opposing field lines are driven together and rapidly reorient into a new configuration, converts magnetic energy into kinetic and thermal energy while also potentially accelerating electrons to high energies (e.g., \citealt{SironiSpitkovsky2014}).
These high energy electrons can radiate at higher electromagnetic frequencies than the bulk of the accretion disk emission and significantly alter the spectral energy distribution (SED, \citealt{Gutierrez2024}), e.g., with a non-thermal component.
The collision of jets with toroidally dominant magnetic field is analogous to collisions of flux tubes in the solar corona, a process that has been proposed as an explanation for certain types of solar flares \citep{Sturrock1984,Hanaoka1994,Falewicz1999,Linton2001}. 

In order to evaluate these and other hypothetical emission mechanisms, it is important to understand the possible accretion states of SMBH binary systems.
For instance, magnetically arrested disk (MAD) accretion flows \citep{Narayan2003,Igumenshchev2003,Sasha2011} have been widely studied in single black hole systems (e.g., \citealt{Narayan2012,Ripperda2020,Ripperda2022,Chatterjee2022}) and are the favored models for Event Horizon Telescope targets M87* \citep{Chael2019,EHT5} and Sagittarius A* \citep{EHT_SGRA_5,Ressler2020,Ressler2023}, as well as possibly most AGN with observable jets \citep{Zamaninasab2014,Nemmen2015,Liska2022,Li2024}.
The MAD state is formed when enough net magnetic flux is accreted onto a black hole to partially inhibit accretion, leading to quasi-periodic cycles of flux accumulation and ejection that may be the source of near-infrared and X-ray flares (e.g., \citealt{Dexter2020b,Porth2021,Ripperda2022}).
MAD accretion flows are associated with the most powerful jets, with Poynting efficiencies (measured with respect to accretion power) that can be $\gtrsim$ 100\% when the black hole is rapidly spinning \citep{Sasha2011}.
In contrast, so-called Standard and Normal Evolution (SANE) flows display weaker jets (e.g., \citealt{Penna2013}) and more small-scale turbulent variability. 
From the small set of magnetohydrodynamic (MHD) simulations of binary accretion to date (e.g., \citealt{Noble2012,Shi2012, Paschalidis2021,Combi2022}), the MAD state in binaries
has only been studied in recent work \citep{Ressler2024, Most2024}.

Moreover, there are only a few general relativistic simulations of binary black hole accretion in the literature, including both force-free electrodynamic simulations \citep{Palenzuela:2009yr,Palenzuela:2010nf,Palenzuela:2009hx,Moesta:2011bn,Alic:2012df} and general relativistic
magnetohydrodynamic (GRMHD) simulations.
This is important because general relativity is required in order to realistically capture the near-horizon flows that are likely the dominant source of emission.
Previous GRMHD simulations can generally be divided into two categories.
The first has focused on accreting disk-like structures \citep{Farris2012, Paschalidis2021,Combi2021,Lopez2021,Avara2024}, while the second has focused on a more uniform distribution of low angular momentum gas \citep{Giacomazzo2012,Kelly2017,Cattorini2021a,Fedrigo2024}.
Simulations that fall into the latter category have been limited to binaries with initial separation distances of $\lesssim 16 M$, where $M$ is the total mass of the two black holes. 
This means that the gas and magnetic field had only a short amount of time to evolve before merger; as a consequence, no clear jet or outflow was observed.
Furthermore, the effective Bondi radii of the gas in these simulations were quite small, $\lesssim 10 \rg$ (here $\rg = GM/c^2$ is the gravitational radius corresponding to the total mass, $M$, of the system, where $G$ is the gravitational constant and $c$ is the speed of light), meaning that the dynamical ranges of accretion that could be studied were restricted. 

In order to limit free parameters and isolate key physics and emission mechanisms, here we focus on the low angular momentum accretion flow scenario (which may be realistic for the large-scale feeding in some lower-luminosity galactic centers, e.g., \citealt{Ressler2018,Ressler2020}, or if the binary decouples from a larger circumbinary disk, e.g., \citealt{Most2024b}).  
Unlike previous work on low angular momentum SMBH binary flows, however, we study gas with significantly larger Bondi radii (150--500$\rg$) and the largest binary separation distances to date (25--27$\rg$). 
This allows us to run our simulations for much longer, $\sim$ 4--6 $\times 10^4$ $\rg/c$ or 70--100 orbits (compared with $\sim$ a few $10^3\, \rg/c$ or $\lesssim$ 10 orbits in previous works).
We do this in full GRMHD using our new implementation of the semi-analytic super-imposed Kerr-Schild metric \citep{Combi2021,Ressler2024,CombiRessler2024} in {\tt Athena++} \citep{White2016,Athenapp}.

In this work, we propose a new flaring mechanism that could potentially result in non-thermal emission involving reconnection caused by ``magnetic bridges'' (connected flux tubes) forming between the two in-spiralling black holes. 
Heating/energization of the plasma that could result in flaring happens when magnetic field lines accrete onto both black holes, get twisted by the orbital motion, and ultimately break off due to magnetic reconnection.
This process effectively extracts orbital energy in the form of thermal and kinetic energy while also likely being a source of high energy particle energization.
The mechanism is analogous to coronal mass ejections in the solar corona \citep{Chen2011}, and has also been proposed for neutron star-neutron star in-spirals \citep{Piro2012,Most:2020ami,Most2022}, neutron star-black hole in-spirals \citep{McWilliams2011,Carrasco2021,Most2023}, and other stellar binary systems (\citealt{Lai2012,Cherkis2021}; see also earlier work on planetary magnetospheres, \citealt{Goldreich1969}).
We also present the first demonstration of jet-jet interactions in GRMHD simulations.

This letter is organized as follows.  
\S \ref{sec:methods} describes our numerical set-up, \S \ref{sec:results} presents our results, while \S \ref{sec:disc_conc} synthesizes these results and concludes. Unless stated otherwise, we adopt units of $G=c=1$.

\section{Methods}
\label{sec:methods}
To simulate the accretion flow onto binary black holes we use a version of the GRMHD portion of {\tt Athena++} \citep{White2016,Athenapp} that has time-dependent metric capabilities as described in \citet{Ressler2024}.  
For the metric itself, we use the superimposed Kerr-Schild metric described in \citet{Combi2021} and \citet{CombiRessler2024} that is constructed by superimposing two linearly boosted Kerr black holes in Cartesian Kerr-Schild coordinates.
Note that we do not include the final temporal interpolation to the remnant black hole and thus limit our study to the in-spiral phase leading up to merger.
The orbits of the black holes in the simulations are calculated from the post-Newtonian (PN) orbital equations using {\tt CBwaves} \citep{cbwaves}, assuming initially circular orbits, with initial separation distances equal to $d_{\rm BH,0}$.  
The angular momentum of the initial orbit points in the $+z$ direction, with black hole dimensionless spin three-vectors of $\bm{\chi_1}$ and $\bm{\chi_2}$. 

We initialize the domain with uniform mass density and pressure parameterized by the Bondi radius $r_{\rm B} = 2M/c_{\rm s,0}^2$, where $c_{\rm s,0}$ is the initial sound speed.  
Since the simulations are scale free we set the initial rest-mass density, $\rho_0$, to unity and thus the initial pressure is given by $P_0 = 2 M/(\gamma r_{\rm B})$, where $\gamma$ is the adiabatic index and we have used the non-relativistic expression for the sound speed appropriate for relatively large Bondi radii.
The fluid velocities are set such that the gas is initially at rest (i.e., the bulk Lorentz factor is unity).
The magnetic field is chosen to be uniform in the $+z$ direction with magnitude set such that the initial $\beta=2 P/ (b^\mu b_\mu)=100$, where $b^\mu$ is the comoving magnetic field in Lorentz-Heaviside units (that is, a factor of $1/\sqrt{4{\rm \pi}}$ has been absorbed in $b^\mu)$. Such coherent magnetic field conditions in a low angular momentum flow may be realized if, e.g., the large-scale circumbinary accretion flow itself is magnetically arrested \citep{Most2024} and the binary has decoupled from a larger circumbinary disk \citep{Most2024b}.

In this work we focus on equal mass binaries: $M_1 = M_2 \equiv 0.5 M$, where $M_i$ are the masses of the individual black holes. 
We measure length and time with respect to $M_{\rm tot}\equiv M = 2M_1$, so that $r_{\rm g}  = M = 2M_1$. 

\subsection{Grid Structure and Numerical Details}
\label{sec:numerical_details}
Our simulations cover a box of (1600 $\rg$)$^3$ with a base resolution of 128$^3$ numerical grid points. 
Using adaptive mesh refinement (AMR) we then add 9 extra levels of refinement approximately every factor of 2 in radial distance from each black hole.  
For each black hole this results in a $128^3$ grid being placed within $ -1.56 \rg \le X,Y,Z \le 1.56 \rg$, where $X,Y$, and $Z$ are the black hole rest frame coordinates centered on the black hole, with cell separations of $\Delta x  \approx 0.024\, \rg$ (or $\approx$ 28 cells from the origin to the event horizon).
Since the event horizon is larger for the $\chi=0$ simulation, we use only 8 extra levels of refinement, meaning that the finest level of  $128^3$ cells is places within $ -3.12 \rg \le X,Y,Z \le 3.12 \rg$, with cell separations of $\Delta x  \approx 0.048\,\rg$ (or $\approx$ 21 cells from the origin to the event horizon).

We use ``outflow'' boundary conditions at the edge of the box in each direction.
In other words, we copy each primitive variable into the ghost zones unless the perpendicular component of the velocity is flowing into the grid.  
In that case we set the perpendicular component of the velocity to zero to prevent inflow from the boundary.  

\begin{figure}
\includegraphics[width=0.49\textwidth]{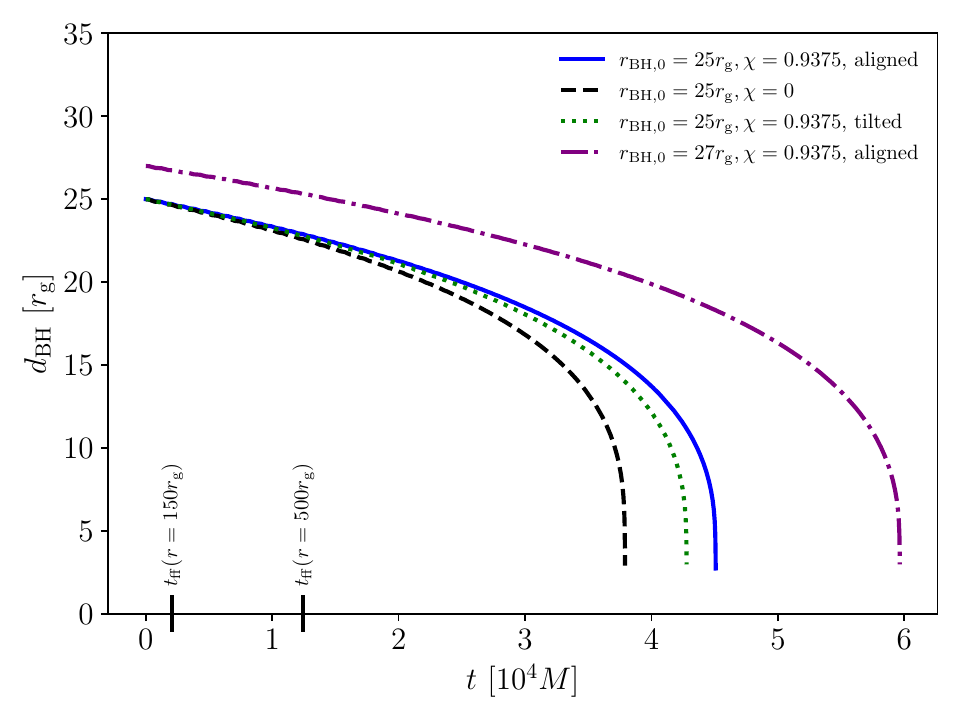}
\caption{Binary separation distance, $d_{\rm BH}$ vs.\ time for our four orbital configurations.
The free fall times at the two choices for Bondi radii, $t_{\rm ff}$, are also shown on the $x$-axis.
The time that it takes for the binaries to merge is a strong function of the initial separation distance, $d_{\rm BH,0}$, with the $d_{\rm BH,0}=27\rg$ orbit taking an extra $\sim$ 2$\times 10^4 M$ than $d_{\rm BH,0}=25\rg$ orbits.
Orbits with nonzero black hole spin aligned with the orbital angular momentum generally take longer to merge due to the additional net angular momentum.}
\label{fig:r_orbits}
\end{figure}

We update the spacetime metric every 10 timesteps for improved numerical efficiency.
Since the black holes are always moving at velocities $\lesssim 0.3c$, the errors incurred by this choice are small when compared with the errors incurred by the GRMHD fluid evolution as argued in \citet{Ressler2024}.

We utilize the same modifications to the fluid and spacetime metric within the event horizons as described in detail in \citet{Ressler2024}.
These modifications do not affect the flow outside the event horizon and ensure numerical stability.

We use the HLLE Riemann solver \citep{Einfeldt1988} and the piece-wise parabolic method \citep{Colella1984} for reconstruction.
The density floor is $10^{-6} (r/r_{\rm g})^{-3/2}$ and the pressure floor is $3.33 \times 10^{-9} (r/r_{\rm g})^{-5/2}$, with $\sigma \equiv b^\mu b_\mu/\rho \le 100$ and $\beta \ge 0.001$ enforced via additional density and pressure floors, respectively.  
Additionally, the velocity of the gas is limited such that the maximum bulk Lorentz factor is $<50$.  

\begin{table*}
\centering
\begin{tabular}{|l|c|c|c|c|c|c|c|c|c|}
\hline
Simulation Name & $d_{\rm BH,0}$ & $r_{\rm B}$ & $\chi$ & $\bm{\chi}_1(t=0)$ & $\bm{\chi}_2(t=0)$ & $t_{\rm merger}$ [$10^4 M$] & $N_{\rm orbits}$ & $T_{\rm orbit}(t=0) [M]$ & $v_{\rm orbit}(t=0)[c]$\\ \hline
Fiducial & $25 \rg$ & $150 \rg$ & 0.9375 & [0,0,$\chi$] &[0,0,$\chi$] & 4.51 & 86 & 834 & 0.094 \\ \hline
$\chi=0$ & $25 \rg$ & $150 \rg$ & 0 &[0,0,0] &[0,0,0] & 3.79 & 70 & 827 & 0.094 \\ \hline
Tilted & $25 \rg$ & $150 \rg$  & 0.9375 & $\frac{\sqrt{2}}{2} \left[\chi,0,\chi\right]$ & $\frac{\sqrt{2}}{2} \left[-\chi,0,\chi\right]$ & 4.28 & 81 & 832 &  0.094\\ \hline
$r_{\rm B}=500\rg$ & $27 \rg$ & $500 \rg$ &  0.9375 & [0,0,$\chi$] & [0,0,$\chi$] & 5.96 &  102 & 932 & 0.091 \\ \hline
\end{tabular}
\caption{Parameters and orbital quantities for our four simulations. Here $d_{\rm BH,0}$ is the initial binary separation distance, $r_{\rm B}$ is the Bondi radius of the gas, $\bm{\chi_i}$ are the spin vectors of the two black holes, $\chi$ is the magnitude of the black hole spins, $t_{\rm merger}$ is the time it takes for the black holes to merge, $N_{\rm orbits}$ is the number of orbits before merger, $T_{\rm orbit}$ is the orbital period, and $v_{\rm orbit}$ is the orbital speed. }
\label{tab:orbits}
\end{table*}

\subsection{Specific Runs}
For this work, we run a total of four simulations, summarized in Table \ref{tab:orbits}.
Three of the simulations use $d_{\rm BH,0}=25 \rg$ and $r_{\rm B} = 150 \rg$.
Of these three, the first uses two rapidly rotating black holes ($\chi_1=\chi_2=0.9375$) aligned with the orbital angular momentum in the $+z$ direction.
The second uses two non-spinning black holes ($\chi_1 =\chi_2 = 0$).
The third uses two rapidly rotating black holes ($\chi_1=\chi_2=0.9375$) both inclined with respect to the $+z$ axis such that initially $\bm{\chi_1} = [\chi_1 \sin(45^\circ),\ 0,\ \chi_1 \cos(45^\circ)] $ and $\bm{\chi_2} = [\chi_2 \sin(-45^\circ),\ 0,\ \chi_2\cos(-45^\circ)]$. 
That is, the spins are initially tilted by $45^\circ$ in opposite directions such that they are perpendicular to each other.  
The fourth simulation uses a significantly larger Bondi radius and a slightly larger binary separation distance, $r_{\rm B} = 500 \rg$ and $d_{\rm BH,0}=27 r_{\rm g}$ with two rapidly rotating black holes ($\chi_1=\chi_2=0.9375$) aligned with the orbital angular momentum in the $+z$ direction.

Our motivation for choosing these particular Bondi radii is primarily computational.  
Realistic values of $r_{\rm B}$ for the centers of galaxies are likely orders of magnitude larger (e.g., \citealt{Garcia2005,Wong2011,Runge2021}, though for SMBH binary-hosting galaxies $r_{\rm B}$ may be moderately lower if the surrounding gas temperatures are higher as a result of the galaxy merger).
Larger Bondi radii, however, require longer simulation run-times in order to reach inflow equilibrium and so studying flows with $r_{\rm B}\gtrsim 1000 \rg$ becomes computationally infeasible in a single simulation.
We are thus limited to studying smaller Bondi radii so that the duration of the simulations are at least several free-fall times at $r_{\rm B}$.
Increasing $r_{\rm B}$ from $150\rg$ to $500\rg$ for one simulation then allows us to investigate how this choice may affect our results. 
We note that even $r_{\rm B}=150\rg$ is significantly larger than the $r_{\rm B}\lesssim 10 \rg$ used in previous GRMHD simulations.

Since all of our simulations use $\chi_1=\chi_2$, in what follows we define $\chi=\chi_1=\chi_2$.
We refer to the $d_{\rm BH,0}=25 r_{\rm g}$, $\chi=0.9375$, $r_{\rm B} = 150 \rg$, aligned run as the `fiducial' simulation.
We refer to the $d_{\rm BH,0}=25 r_{\rm g}$, $\chi=0$, $r_{\rm B} = 150 \rg$ run as the `$\chi=0$' simulation.
We refer to the $d_{\rm BH,0}=25 r_{\rm g}$, $\chi=0.9375$, $r_{\rm B} = 150 \rg$, tilted run as the `tilted' simulation.
Finally, we refer to the $d_{\rm BH,0}=27 r_{\rm g}$, $\chi=0.9375$, $r_{\rm B} = 500 \rg$ run as the `$r_{\rm B} = 500 \rg$' simulation.

\subsection{Binary Orbits}

 \begin{figure*}
\includegraphics[width=0.95\textwidth]
{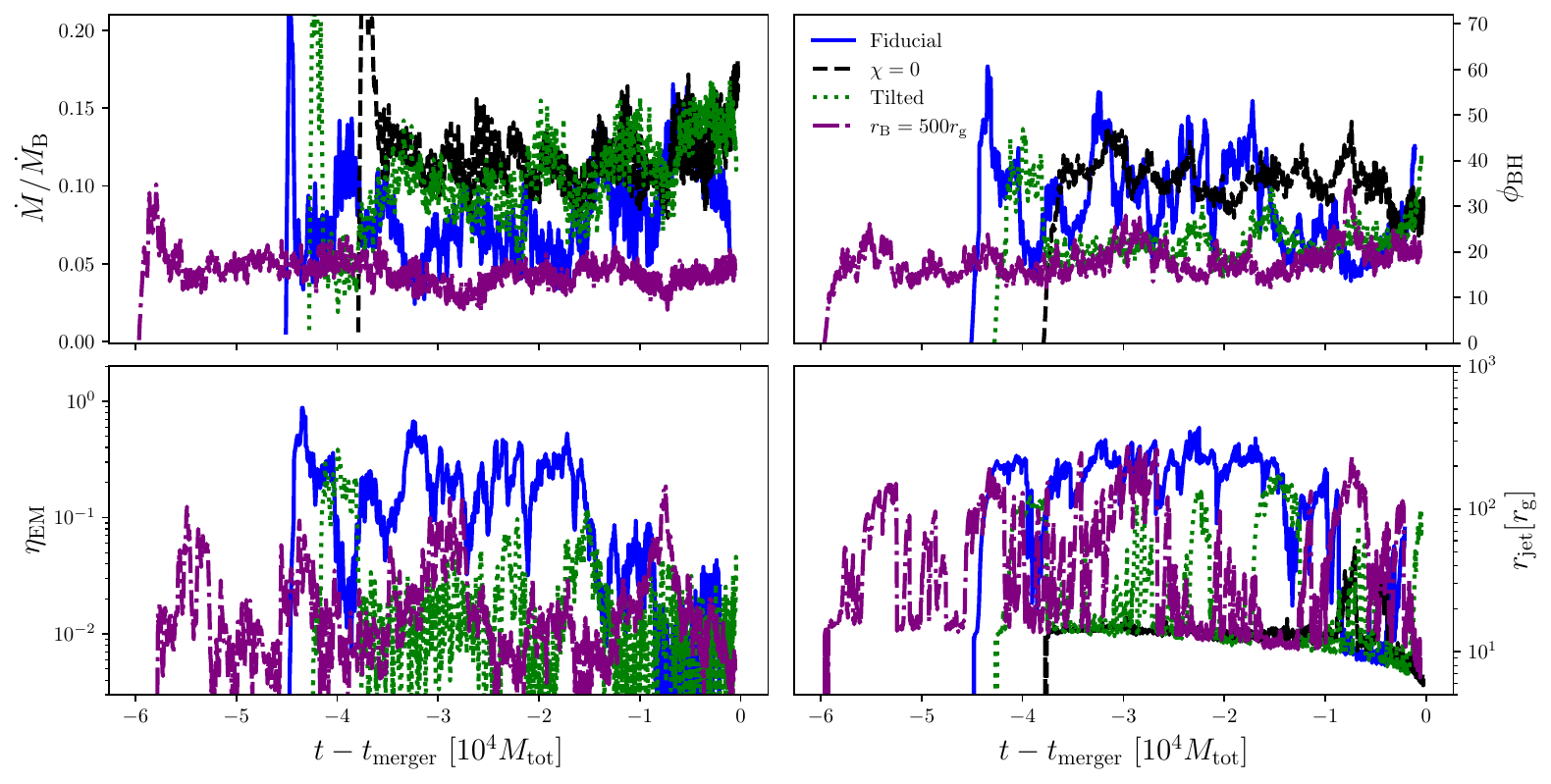}
\caption{Time series of accretion and jet-related quantities.  Top left: total accretion rate, $\dot M$, onto both black holes normalized to the Bondi rate, $\dot M_{\rm B}$. 
Top right: total dimensionless magnetic flux threading both black holes, $\phi_{\rm BH}$ (calculated from Equation \ref{eq:phibh}).
Bottom left: total electromagnetic energy outflow efficiency, $\eta_{\rm EM}$.
Bottom right: maximum distance from the origin reached by the jets, $r_{\rm jet}$, measured as the maximum distance at which gas with $\sigma>1$ is found. 
In calculating $\phi_{\rm BH}$ and $\eta_{\rm EM}$, the accretion rates in the denominator have been averaged over one orbital time. 
All simulations show approximate saturation of magnetic flux at some mean value.
The fiducial and $\chi=0$ simulations saturate at higher horizon-penetrating magnetic fluxes ($\phi_{\rm BH} \sim 30$--$50$) and display strong variability. 
The tilted and $r_{\rm B}=500 \rg$ simulations, on the other hand, saturate at $\phi_{\rm BH}$ $ \approx 20$--$30$, with less variability.
Correspondingly, the jet efficiencies in the fiducial simulation are the largest, with $\eta_{\rm EM} \sim 20-50\%$ on average, while the efficiencies in the tilted and $r_{\rm B}=500 \rg$ simulations only occasionally rise above a few percent.
This shows that both larger Bondi radii and tilted accretion flows can inhibit the formation of strong jets in equal mass black hole binaries.
}
\label{fig:time_plots}
\end{figure*}

The orbital separation distances as calculated from {\tt CBWAVES} for the four configurations are shown as a function of time in Figure \ref{fig:r_orbits}.
For the $d_{\rm BH,0} =25\rg$ separation runs, merger occurs somewhere between 3.8--4.5 $\times 10^4 M$ (or 70--86 orbits), where the shorter time corresponds to nonspinning black holes ($\chi=0$) and the longer time corresponds to aligned $\chi=0.9375$ black holes.
The tilted $\chi=0.9375$ binary falls somewhere in between (note that in this case both spin vectors are positive when projected onto the orbital angular momentum axis, which is aligned with the $z$-direction).
This is because the binaries with nonzero black hole spin have a larger total angular momentum (including orbital angular momentum and black hole spin), which takes longer to shed via gravitational waves.  
The $d_{\rm BH,0}=27\rg$ binary merges on a slightly longer timescale, $\approx 6 \times 10^4 M $ (or 102 orbits), consistent with an $d_{\rm BH,0}^4$ scaling.

These merger timescales can be compared to the free-fall times at the Bondi radii, defined as 
\begin{equation}
    t_{\rm ff} (r) = \frac{\rm \pi}{2 \sqrt{2}} \left(\frac{r}{\rg}\right)^{3/2} M,
\end{equation}
where this expression is valid only for $r$ much greater than the orbital radius because it assumes Newtonian gravity and spherical symmetry.
For $r_{\rm B}=150 \rg$ and $r_{\rm B}=500 \rg$, $t_{\rm ff}$ is $\approx$ 2{,}040 $M$ and $\approx$ 12{,}420 $M$, respectively.
So for $r_{\rm B}=150 \rg$ we can simulate up to $\sim$ 19 free-fall times while for $r_{\rm B}=500 \rg$ we can simulate up to $\sim$ 5 free-fall times before merger.

Note that the merger timescales are also longer than the free-fall timescales at the outer boundary of the simulation ($\approx 
 2.5 \times 10^4 $ $M$) by a factor of $\sim$ 2. 
 However, gas outside of the Bondi radius, by definition, is not strongly affected by the gravity of the black hole and thus does not significantly evolve over the duration of the simulation.
On the other hand, outflows in the form of shocks caused by jets and the orbital motion of the black holes do propagate to the outer boundary.  
The outflow boundary conditions we use (see \S \ref{sec:numerical_details}) do a good job of allowing these outflows to freely propagate through the boundary without artificial reflection.

For the orbital configuration with black hole spins initially tilted with respect to the orbital angular momentum axis (the $z$-axis), the spin directions will precess with time.  
In general, the orbital angular momentum direction would also precess with time.
Because of our choice of to have the net black hole spin $\bm{\chi_1}+\bm{\chi_2}$ aligned with the orbital angular momentum axis, however, the latter is constant with time.
The spins, on the other hand, precess about the $z$-axis in a clockwise fashion such that the net spin, $\bm{\chi_1}+\bm{\chi_2}$, remains constant in both direction and magnitude. 
The period of this precession is initially $\approx$ $2 \times 10^4 M$, much longer than the orbital period, and grows shorter as the binary approaches merger.

\begin{figure*}
\includegraphics[width=0.99\textwidth]
{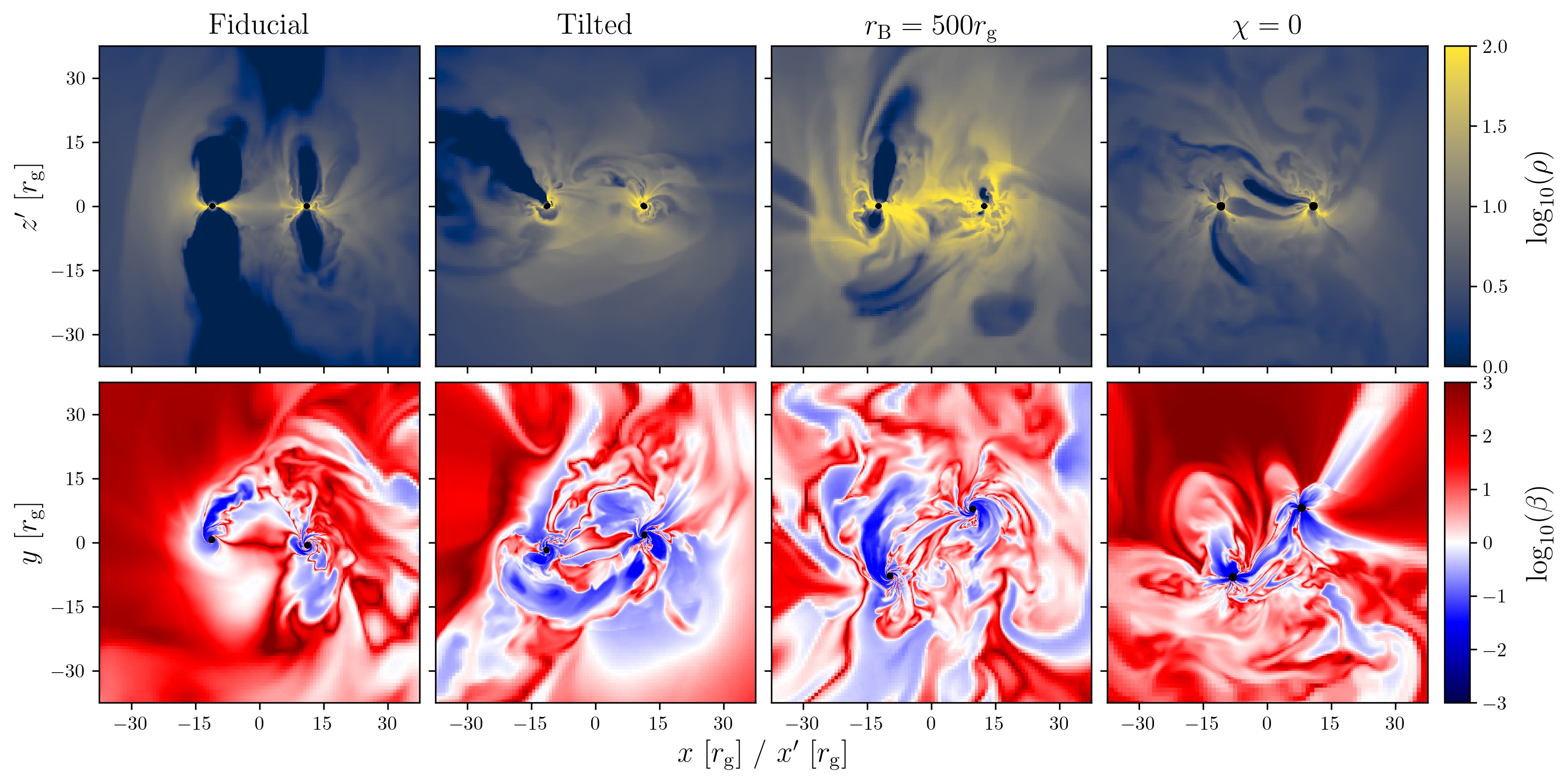}
\caption{Top: Slices of rest-mass density, ${\rho}$, in the co-orbiting frame at representative times in our four simulations. 
Here $x^\prime$ and $z^\prime$ are the coordinates in the co-orbiting frame.
Bottom: Equatorial slices of plasma $\beta$ at representative times in our four simulations.
Only the fiducial simulation shows a clear structure of an accreting midplane and poloidal jet.  
The other simulations show more unstructured accretion flows, with occasional weak and intermittent jets in the tilted and $r_{\rm B}=500 \rg$ simulations.
In the midplane, regions of low $\beta$ are seen propagating outwards from the near-horizon flow and mixing with the gas at larger radii. 
These include both ``flux tubes'' like those seen in MAD accretion flows and bubbles blown out by weak jets (in the case of the tilted simulation).
The low $\beta$ regions are particularly large and mix to particularly large radii in the tilted and $r_{\rm B}=500 \rg$ simulations.
Such mixing likely inhibits the accumulation of net magnetic flux on the event horizon, required for sustained jet launching. 
Animations of this figure are available for the \href{https://youtu.be/XQbifuBWjw8}{fiducial}, \href{https://youtu.be/yoiAEejPz2Q}{$\chi=0$}, \href{https://youtu.be/I5UnUlYVHag}{tilted}, and \href{https://youtu.be/YrTxBuysitk}{$r_{\rm B}=500\rg$} simulations.
}
\label{fig:corotating_slice}
\end{figure*}

\section{Results}
\label{sec:results}
\subsection{General Flow Properties}
Figure \ref{fig:time_plots} shows the time evolution of the total mass accretion rate (i.e., the sum of the accretion rate onto each black hole) normalized to the Bondi rate, $(|\dot M_1+\dot M_2|)/ \dot M_{\rm B} \equiv |\dot M|/\dot M_{\rm B}$, the normalized total magnetic flux threading the two black holes, 
\begin{equation}
    \label{eq:phibh}
    \phi_{\rm BH} = \frac{\Phi_{\rm BH,1}+\Phi_{\rm BH,2}}{\sqrt{|\dot M_1|}+\sqrt{|\dot M_2|}},
\end{equation} 
the total electromagnetic outflow efficiency,
\begin{equation}
    \label{eq:etaem}
    \eta_{\rm EM} = \frac{\dot E_{\rm EM,1}+\dot E_{\rm EM,2}}{|\dot M_1|+|\dot M_2|},
\end{equation} 
and the maximum distance from the origin reached by the jets, $r_{\rm jet}$, in our four simulations.  
Here $\dot M_{\rm B}$ is calculated from the initial conditions using the total mass of the two black holes, $\Phi_{\rm BH,i}$ is calculated by first transforming the three-magnetic field, $B^i$, to locally spherical coordinates in each black hole's rest frame, then integrating $|B^r|$ over the surface of the horizon, $\dot E_{\rm EM,i}$ is the electromagnetic energy outflow rate for each black hole measured just outside the event horizon, and $r_{\rm jet}$ is defined as the maximum radius at which there is any gas with $\sigma >1$.

Both the fiducial and $\chi=0$ simulations saturate at relatively high values of $\phi_{\rm BH}$, $\sim$ 40--50, displaying variability characteristic of the magnetically arrested state (namely, there are cycles of flux accumulation followed by dissipation\footnote{The flux accumulation and dissipation cycles are clearer when $\Phi_{\rm BH,i} $ (the unnormalized magnetic flux threading the event horizon) is plotted for one of the black holes.  We plot the normalized $\phi_{\rm BH}$ instead because its magnitude is typically used to classify the magnetically arrested state.} centered on a saturated mean value). 
Dissipation events do not occur simultaneously for each black hole; this tends to wash out the characteristic MAD variability in the total $\varphi_{\rm BH}$ when compared to an individual black hole's $\varphi_{\rm BH,i}$.
Note that we avoid precisely classifying our simulations as ``magnetically arrested'' due to the ambiguity of the term as we discuss in Appendix \ref{app:MAD}.
Both the tilted and $r_{\rm B} = 500 \rg$ simulations display lower values of $\phi_{\rm BH}$ than the other two simulations, $\sim$ 20--30, though they also saturate around a mean value and show some hints of flux accumulation and dissipation. 
Only the fiducial simulation shows moderately efficient (though variable) jets, with $\eta_{\rm EM}$ typically between 20--50\%. 
The other simulations show efficiencies generally at the $\sim$ few percent level with occasional transient periods of efficiencies $\lesssim 10 \%$.
Because of the increased outflow, the fiducial simulation also displays lower accretion rates than the tilted and $\chi=0$ simulations, $\dot M/\dot M_{\rm B}$ $\sim$ 8\% compared to $\sim$ 12\%.
The $r_{\rm B} = 500 \rg$ simulation has an even lower accretion rate as expected due to the larger initial Bondi radius, with $\dot M/\dot M_{\rm B}$ $\sim$ 5\%.
Specifically, this is caused by the radial dependence of the mass density with radius, which we find to be $\tilde \propto$ $r^{-1}$, consistent with the general result for hot turbulent flows around black holes \citep{Pen2003,Ressler2021,Ressler2023,SCAF}.
Larger Bondi radius flows then have lower accretion rates relative to the Bondi rate by a factor of $\tilde \propto$ $r_{\rm B}^{-1/2}$ (or $\sim$ 0.5 for the specific case of $r_{\rm B}$=500 $\rg$ vs.\ $r_{\rm B}$=150 $\rg$).

To understand why the fiducial and $\chi=0$ simulations saturate at higher values of $\phi_{\rm BH}$ than the tilted and $r_{\rm B} = 500\rg$ simulations despite similar initial conditions, in Figure \ref{fig:corotating_slice} we plot slices of mass density in the corotating frame $x^\prime$--$z^\prime$ and slices of plasma $\beta$ in the $x$-$y$ plane for each simulation.
Only the fiducial simulation shows a clear structure of an accreting midplane and low density polar outflows for each black hole.  
The other simulations show more complicated structures, with low density polar regions associated with outflows only present intermittently.
In the midplane, all simulations show ``flux tubes'' of highly magnetized regions buoyantly rising to larger radii.  
These flux tubes are a result of flux ejection near the event horizon, typical for flows showing signs of being magnetically arrested (e.g., \citealt{Sasha2011}).
They contain predominantly vertical magnetic fields and are formed by reconnection of midplane current sheets (see \citealt{Ripperda2022} for more details on this mechanism).
The flux tubes are ejected due to magnetic field lines accreting with the radially infalling gas that subsequently pile up near the horizon, form a thin layer due to the build-up of magnetic pressure, and ultimately reconnect.  
For the tilted simulation, weak jets are also sometimes seen in the midplane; these weak jets appear
similar to larger flux tubes.  
Comparing the four simulations, the tilted and $r_{\rm B}=500 \rg$ simulations show the largest regions of low $\beta$; these regions also extend to larger radii compared with those in the fiducial and $\chi=0$ simulations.  
In the tilted case, this is because of the misaligned (though weak) jets interacting with the midplane accretion flow.
In the $r_{\rm B}=500 \rg$ case, it is caused by the larger dynamical range of the flow resulting in an extended region of turbulence that affects the amount of magnetic flux that can reach the horizon coming from larger scales.  
More precisely, since field lines are being accreted from a larger initial radius, the (near-equatorial) current sheets that form near the horizon when the flux accumulates are also longer, naturally leading to larger flux tubes\footnote{In general, longer current sheets are associated with longer field lines that can be transformed into flux tubes.
Moreover, since the reconnection rate is independent of current sheet geometry, longer current sheets spend longer amounts of time reconnecting, resulting in a larger amount of reconnected flux.  
As a result, flux tube size tends to increase with the increasing current sheet length.
This has been shown for standard MADs in \citet{Ripperda2022} and also for initially zero angular momentum flows in \citet{Galishnikova2024}.} that get propelled to larger distances.
As these flux tubes propagate outwards they tangle the geometry of the inflowing magnetic field while also removing vertical magnetic flux from the near-horizon region.
In both the tilted and $r_{\rm B}=500 \rg$ simulations, therefore, feedback in the form of strongly magnetized regions alters the flow enough to reduce the net magnetic flux reaching the event horizon and prevent strong jets from forming. 

We note, however, that all of our simulations show this ``flux tube feedback''; the effect is just stronger in the tilted and $r_{\rm B} = 500\rg$ simulations.
This is evidenced by the saturation of magnetic flux for all simulations in Figure \ref{fig:time_plots} and by the noticeable flux tubes ejected in the midplane seen in Figure \ref{fig:corotating_slice}, both of which are reminiscent of magnetically arrested flows. 
Whether or not our simulations can be truly classified as magnetically arrested is partly semantic, as we argue in Appendix \ref{app:MAD}.
Generally, flux tubes are ejected in the direction opposite of the black hole motion (see bottom row of Figure \ref{fig:corotating_slice}, where the black holes are moving in a counter-clockwise direction).
This is because the current sheets that form from accreted poloidal magnetic field are preferentially formed ``behind'' the black holes as they drag the accreted field lines (see, e.g., \citealt{Palenzuela2010,Neilsen2011,Most2023} for binaries, and \citealt{Penna2015,Cayuso2019,Kim:2024zjb} for boosted isolated black holes).
As a result, the current sheets are larger (i.e., have more reconnecting surface) and magnetic reconnection dissipates more energy than for stationary black holes, all else being equal.
Such enhanced dissipation may be the reason that even our fiducial and $\chi=0$ simulations do not reach the maximal values of $\phi_{\rm BH}$ (50--70) or jet efficiency ($>100\%$) seen in ``traditional'' MAD flows (e.g., \citealt{Sasha2011,Narayan2012,Ressler2021,Chatterjee2022,Ressler2023,Lalakos2024}).
Nevertheless, precisely quantifying the effect of black hole orbital motion on the feedback-regulated accretion of magnetic flux and disentangling that from fluid effects is difficult without a larger parameter survey.
If indeed orbital motion is enough to prevent maximally efficient jets from forming, this would have important consequences for jet feedback in and observations of SMBH binary AGN.

It is instructive to compare our findings to single black hole accretion simulations.
Our result that larger Bondi radius flows heave weaker jets and lower horizon-penetrating magnetic flux is consistent with recent findings in single black hole accretion for low angular momentum flows \citep{Lalakos2024,Galishnikova2024,Kim:2024zjb}, where it is argued that both jet and flux tube feedback tangle the incoming magnetic field.
On the other hand, simulations of tilted accretion flows around single black holes have found that only highly misaligned ($\gtrsim$ 60\degree) flows result in lower horizon-penetrating magnetic flux and weaker jets \citep{Ressler2023,Chatterjee2023}.
The difference here is that the misalignment is with respect to the orbital plane of the binary, not the angular momentum of the gas.  
For an isolated black hole, rapid spin and strong magnetic fields can align the accretion flow and jet with the spin axis as long as the tilt angles are not extreme ($\gtrsim 60 \degree$).
Conversely, in binary accretion flows, the orbital angular momentum is either unchanged by black hole spin (if $\bm{\chi_1}+\bm{\chi_2}$ is constant) or changes on long time scales (if $\bm{\chi_1}+\bm{\chi_2}$ varies due to spin-orbit coupling, see, e.g., \citealt{Ressler2024}).
This means that any jet (even weak jets) will often collide with the incoming accretion flow in the orbital plane and partially inhibit the accretion of net magnetic flux.
It is unclear whether the black hole spins of SMBH binaries in the in-spiral and merger phases are expected to be aligned, as it depends on the accretion and dynamical history of the two black holes leading up to the gravitational-wave emitting regime.
If misaligned spins are common, SMBH binary AGN with strong jets may be rarer than their single black hole AGN counterparts.
It will be important to study a larger range of parameter space (including circumbinary disk accretion) in order to investigate the generality of this finding.

\begin{figure*}
\includegraphics[width=0.99\textwidth]
{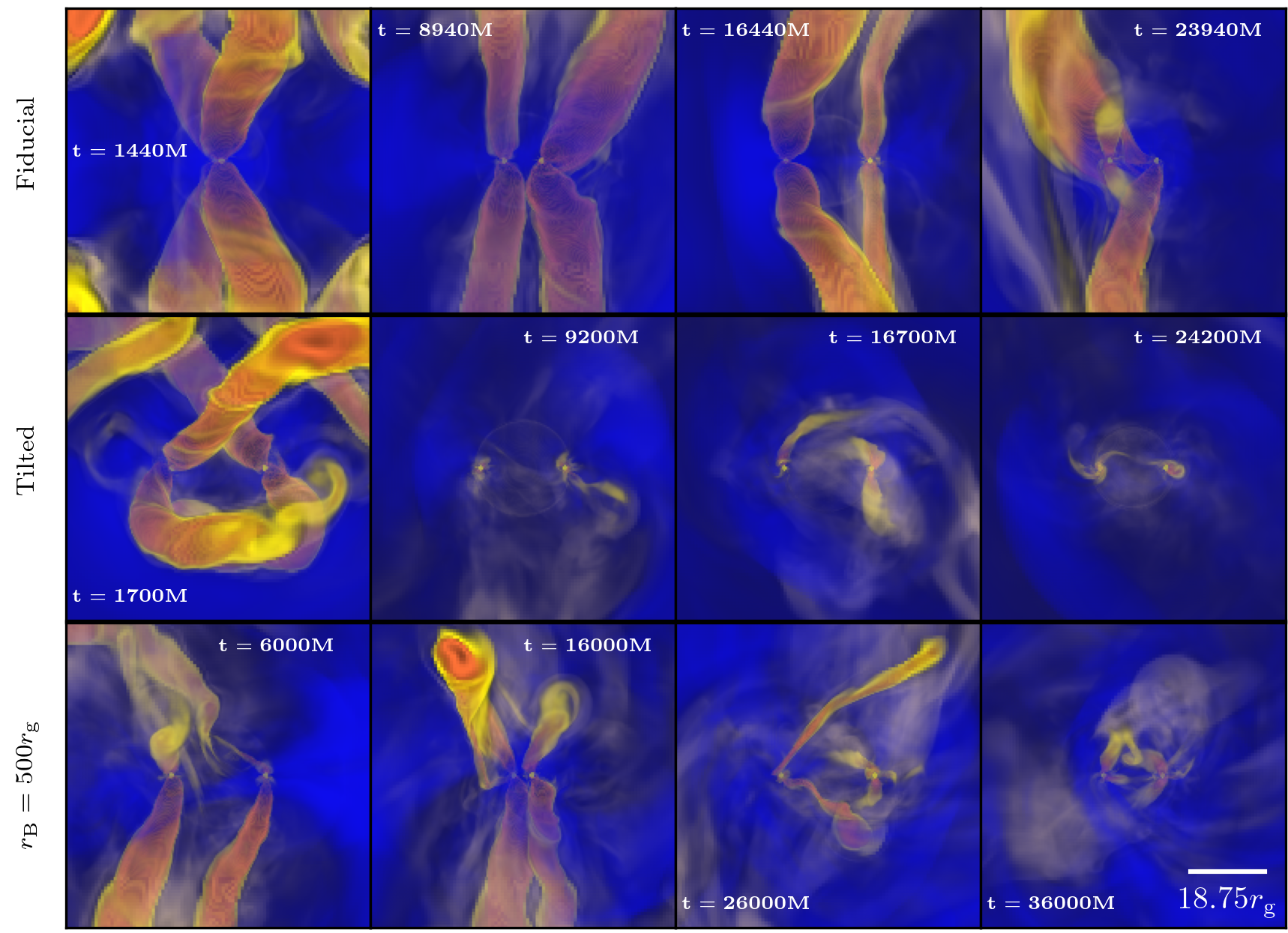}
\caption{ 
Three-dimensional volume renderings of our fiducial (top row), tilted (middle row) and $r_{\rm B}=500 \rg$ (bottom row) simulations at four different representative times, 
highlighting regions of high magnetization, $\sigma$, in orange/yellow and regions of high $\rho$ in blue.  
Each panel is on a scale of $(75 \rg)^2$.
Jets in the fiducial simulation are much more sustained and propagate more effectively than in both the tilted and $r_{\rm B}=500 \rg$ simulations.  
In the latter two simulations the jets are sporadic and intermittent.  
This result may imply that sustained jets are more difficult to form in SMBH binary systems when the spins are misaligned with the orbital plane. 
Whether or not this result holds for different types of large-scale accretion flows (e.g., circumbinary disks) requires future study.
Animations of this figure are available for the \href{https://youtu.be/rMieXN7HFvg}{fiducial}, \href{https://youtu.be/U6pGvOZ83Zo}{tilted}, and \href{https://youtu.be/-l8sEZXR2zg}{$r_{\rm B}=500\rg$} simulations.
}
\label{fig:3D_jet}
\end{figure*}

Another important finding shown in Figure \ref{fig:time_plots} is that $\phi_{\rm BH}$ in both the fiducial and $\chi=0$ simulations decrease shortly before merger (within $\sim$ 5 $\times 10^3 M$).  
This is particularly evident in the fiducial simulation where the jet power and radius drop substantially at that time (with $\phi_{\rm BH}$ dipping to $\sim$ 30, comparable to the tilted and $r_{\rm B} = 500 \rg$ simulations).
The reason for this is that at these late times the black holes' orbital velocities start increasing rapidly, changing the geometry of the accretion flow and reducing the amount of net vertical flux being accreted.

\subsection{Jet Propagation }

Figure \ref{fig:3D_jet} shows volume renderings of the three simulations with nonzero black hole spins at four different representative times on a $(75 \rg)^2$ scale. 
Regions with high $\sigma$ are highlighted in orange/yellow and high density regions are highlighted in blue.  
Jets in the fiducial simulation are persistent, propagate to large radii, and spiral around each other as the black holes orbit.
Jets in the tilted simulation start out clear and structured, following the black hole spin axes and colliding once an orbit.  
As time proceeds, however, the jets die off and only very rarely appear structured.  
They also only rarely reach any significant distance from the black holes (see also the bottom right panel of Figure \ref{fig:time_plots}).
This is likely caused by feedback from the misaligned black holes inhibiting the infall of magnetic flux, limiting the jets' power.
Jets in the $r_{\rm B}=500 \rg$ simulation never have a clear or consistent structure, though they do occasionally reach to relatively large radii (see also the bottom right panel of Figure \ref{fig:time_plots}).
They also often propagate at semi-random angles with respect to the aligned black hole spins.
This is likely caused by the increased amount of turbulence in the accretion flow as discussed in the previous subsection.

Because all simulations contain a significant amount of low angular momentum gas above the orbital plane, all of the jets are also subject to the kink instability.  
This instability occurs when jets with toroidally dominant magnetic fields push against infalling or stationary gas \citep{Begelman1998,Lyubarskii1999}, causing them to wobble or even completely disrupt in extreme cases.  
For a low angular momentum density distribution of $\rho$  $\tilde \propto$  $r^{-1}$ as we find in our simulations (see previous subsection), disruption by the kink instability is likely inevitable \citep{Bromberg2011,Bromberg2016,Sasha2016,Ressler2021,Lalakos2024}.
This may be why all of the jets in our simulations stall by distances of $\sim 300 \rg$ (bottom right panel of Figure \ref{fig:time_plots}). 
On the other hand, at these distances 1) the narrow jets become difficult to resolve without specifically targeted mesh refinement and thus may be subject to larger numerical dissipation and 2) the jets may be affected by the boundary of the simulation, located at a distance of $\sim 800 \rg$.  
While it is important to understand this instability for observations of jets and their emission, a detailed analysis is beyond the scope and focus of this work.

\subsection{Jet Collisions and Dissipation}

\begin{figure}
\includegraphics[width=0.49\textwidth]
{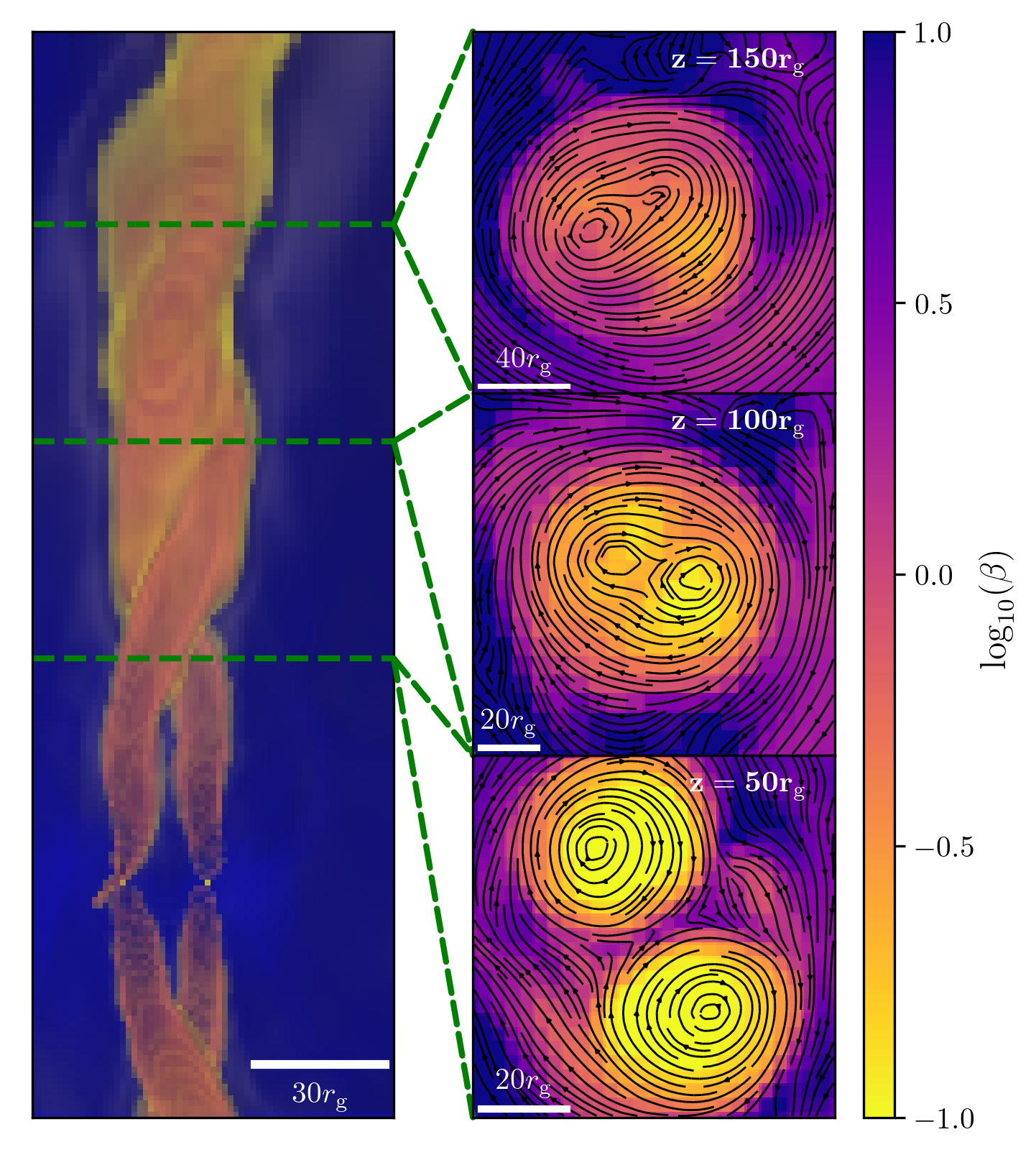}
\caption{
Evidence of magnetic reconnection between two jets in our fiducial simulation.
Left: three-dimensional rendering of the jets at a representative time.  
Right: slices of plasma parameter $\beta$ through the $z=50 \rg$ (bottom panel), $z=100 \rg$ (middle panel), and $z=150 \rg$ (top panel) planes.
Since the spins of the black holes have the same polarity, the in-plane magnetic fields in the jet cores are usually oriented in clockwise loops.  
Because of this, `$X$'-points are formed when the jets initially touch, with oppositely directed field lines driven to reconnect and dissipate.
This is seen as the jet propagates to larger radii and the cores merge into one.
This reconnection layer can accelerate electrons to nonthermal energies which could radiate at high electromagnetic frequencies (as estimated in the main text and Appendix \ref{app:recon_jet}).
An animated version of this figure is available \href{https://youtu.be/KI5tQbRJ9k4}{here}.
}
\label{fig:jet_interaction}
\end{figure}

Focusing now on the fiducial simulation with consistent jets, we show slices through the jet cores of plasma $\beta$ with magnetic field lines overplotted at three different distances from the black holes in Figure \ref{fig:jet_interaction}.
These slices are shown next to a three-dimensional rendering of the jet and are taken at a representative time.
The two \citet{BZ1977} jets tend to have the same polarity of magnetic field because the black hole spins are aligned and the accreted vertical magnetic field is typically of the same sign for both black holes. 
Because of this, when the jet cores first touch at $z\approx 50 \rg$ the magnetic field lines are pointed in opposite directions at the contact surface.  
This results in magnetic reconnection as evidenced by the presence of an `$X$' point, reminiscent of the coalescence of two flux tubes \citep{Lyutikov2017,Ripperda2019a}.  
The reconnection occurs as the jets propagate outwards and continue to be driven together.
Ultimately (by $z \gtrsim 150 \rg$), the jet cores fully merge into a single clockwise loop of magnetic field lines, similar to what has been observed in force-free simulations \citep{Palenzuela:2010nf}.
This process of consistent magnetic reconnection not only dissipates magnetic energy and converts it into kinetic and thermal energy but could also be a source of high energy particle acceleration.

Measuring the properties upstream of the $X$-point at $z=50\rg$ in Appendix \ref{app:recon_jet}, we find that $\sigma$ and $\beta$ are essentially at their maximum and minimum values, respectively (which are set by enforcing a mass density/pressure floor).  
In reality, therefore, $\sigma$ would be much higher and $\beta$ would be much lower since the mass density would likely be much lower, and dominated by pair production and mass-loading of the jet, two physical processes not captured here.  
We also find that the reconnection layer has a significant guide field (that is, a non-reconnecting field perpendicular to the reconnecting plane), with the magnitude of $b^z$ measured in a frame co-moving with the jets (i.e., the poloidal field) roughly equal to the magnitude of the in-plane reconnecting field (i.e., the toroidal field of the jet) as measured in the same frame.

The ratio between the magnitude of the guide field and the magnitude of the reconnecting field is known to be an important parameter in studies of magnetic reconnection, with large values tending to lower the amount of magnetic energy dissipation and the maximum energy of accelerated particles while steepening the nonthermal particle energy spectrum \citep{Zenitani2008,SironiSpitkovsky2012}.
Precisely quantifying this effect in three-dimensional simulations for the parameter regime relevant to this work is still an active area of research.
However, for a guide field of comparable strength to the reconnecting field as we have here, some recent work has found that this effect may be only moderate in three-dimensional simulations, with the energy gain of electrons in one case reduced by $\lesssim 20 \%$ compared with weak guide field reconnection \citep{Werner2024}, and in another case a maximum particle energy reduced by a factor of $\sim$ 2, with a steepening of the nonthermal power law spectral index by $\sim$ 1.5 to $\sim$ 3 \citep{Hoshino2024}.
Note that we quote these numbers only as preliminary estimates, further study of three-dimensional guide field reconnection in the precise parameter regime appropriate for colliding jets is needed. 

In order to obtain more insight into how the jet-jet reconnection layer might appear electromagnetically, in Appendix \ref{app:recon_jet} we estimate the energy released by reconnection between the two jets as 
\begin{equation}
        \frac{\left( L_{\rm rec}\right)_{\rm jet}}{L} \approx\ 0.02\   \left(\left. \frac{u_{b,\rm rec} r_{\rm g}^2}{|\dot M |c} \right/ 5 \times 10^{-5}\right) \left(\frac{l_{\rm rec}}{ 20 \rg}\right)^2,
\end{equation}
where $L$ is the total thermal luminosity of the accretion flow. $u_{\rm b,rec}$ is the magnetic energy density of the reconnecting field, and $l_{\rm rec}$ is the length of the current sheet.  
Alternatively we can write this as
\begin{equation}
\begin{aligned}
        \left( L_{\rm rec}\right)_{\rm jet} \approx\ & 3\times 10^{42} \frac{\mathrm{erg}}{\mathrm{s}}  \left(\left. \frac{u_{b,\rm rec} r_{\rm g}^2}{|\dot M |c} \right/ 5 \times 10^{-5}\right) \left(\frac{l_{\rm rec}}{ 20 \rg}\right)^2 \\
        &\times f_{\rm Edd}\left(\frac{M}{10^6 M_\odot}\right) \left(\frac{\eta_{\rm rad}}{0.1}\right)^{-1},
\end{aligned}
\end{equation}
where $f_{\rm Edd}$ is the Eddington ratio of the source and $\eta_{\rm rad}$ is the radiative efficiency of the flow.
We further estimate that the reconnection happens in the radiative regime and that this energy would be emitted at the so-called ``synchrotron burn-off limit'' of 
\begin{equation}
(\nu_{\rm synch})_{\rm jet} \approx 2.5 \times 10^{22}\ \mathrm{Hz},
\end{equation}
independent of other parameters.  
This frequency is much higher than the estimated thermal synchrotron peak frequency (by at least a factor of $>10^4$, see Appendix \ref{app:recon_therm}) and thus would outshine the thermal emission at this frequency.

We note that the formation of a reconnection layer between the jets crucially depends on the magnetic polarity of the jets being the same.
For jets with opposite polarity (which would form, e.g., when the $z$-component of the spins are anti-aligned), instead of forming a reconnection layer between them the jets may instead bounce off of each other \citep{Linton2001}.
This bouncing may induce a ``tilt-kink'' instability where reconnection happens on the outer layers of the jets \citep{Ripperda2017a,Ripperda2017b}.  
If strong jets in misaligned flows are possible (see discussion in previous subsection), then various other intermediate geometries are possible, subject to more or less magnetic reconnection and higher levels of quasi-periodic variability.  
Future work should explore the broader range of parameter space.

\subsection{Magnetic Bridges}

For simulations without persistent jets, we also find evidence of another type of flaring event.
During these events, what we call ``magnetic bridges'' (connected flux tubes) form between the two black holes that can twist magnetic fields, erupt, and drive hot outflows via reconnection analogous to scenarios proposed for interacting neutron star magnetospheres \citep{Most:2020ami,Most2023}. 

\begin{figure*}
\includegraphics[width=0.99\textwidth]
{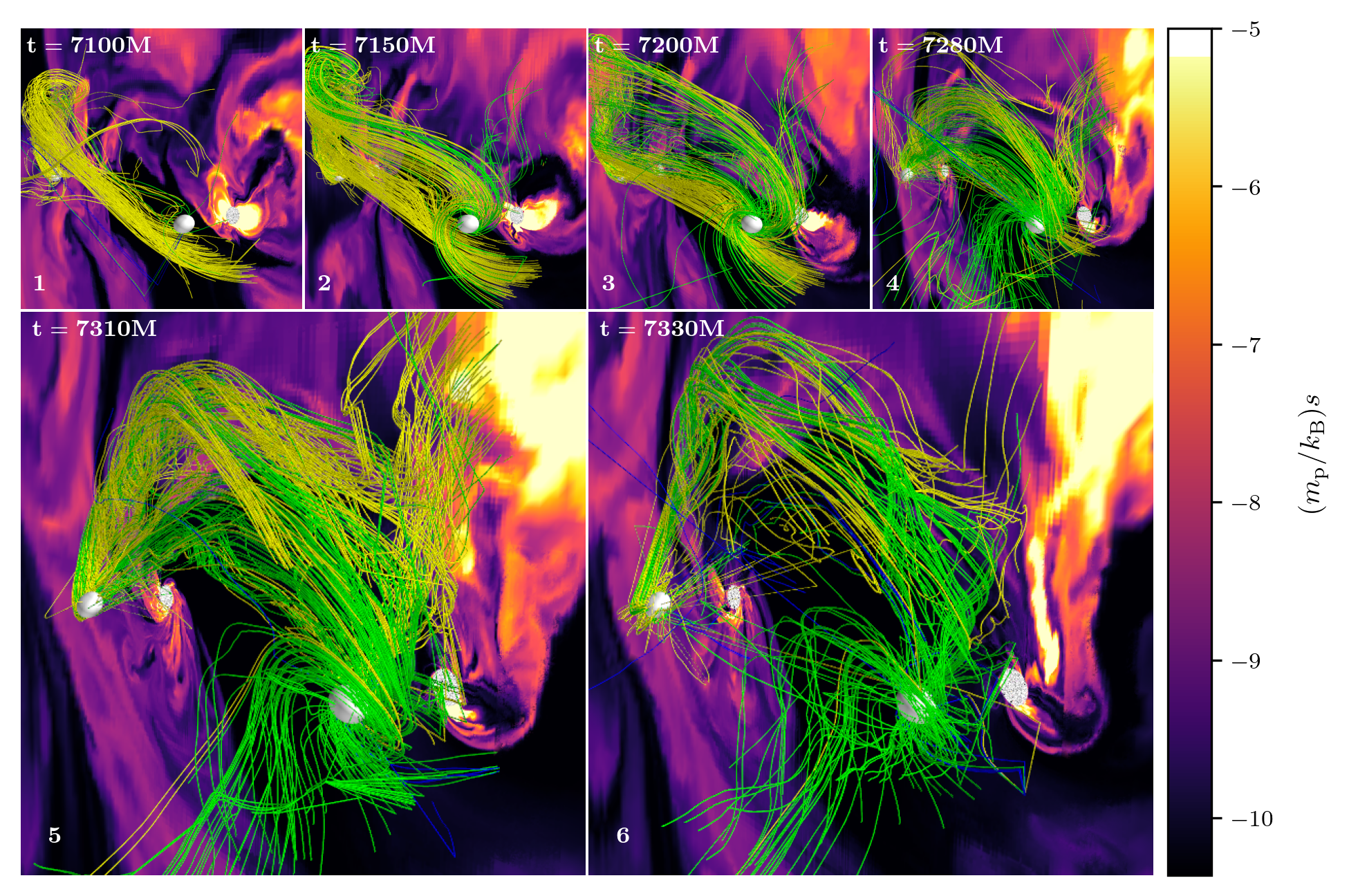}
\caption{
An example of a ``magnetic bridge'' flaring event in our tilted simulation.
Shown in Panels 1--6 is a time series of three-dimensional representations of magnetic field lines in the corotating frame on top of a two-dimensional contour of entropy per unit mass, $s = 1/(\gamma-1) \log (P/\rho^\gamma)$, in the $x^\prime$-$z^\prime$ plane. 
Only field lines that thread at least one of the black hole event horizons \emph{and} get closer than 10 $\rg$ to the other black hole are shown.
Yellow (blue) field lines are those which pierce the first (second) black hole's event horizon (the first black hole is always positioned on the left in these plots) while green field lines are those which pierce both event horizons.
The gray shaded regions in the entropy contour represent the regions inside the event horizons of the black holes.
The second black hole captures initially fairly straight field lines connected to the first black hole and through a combination of orbital and black hole spin motion twists them into a tangled tube.
Finally, the field lines reconnect close to the second black hole, releasing magnetic energy and causing an eruption of high entropy gas that gets expelled to larger radii.
This mechanism is a promising candidate for unique electromagnetically powered high-energy flaring signatures of SMBH binary AGN.
An animation of this figure is available \href{https://youtu.be/6xs0QUaJX5s}{here}.
}
\label{fig:3D_fieldlines}
\end{figure*}

One way of identifying potential flaring mechanisms in GRMHD simulations is through their heating properties (e.g., \citealt{Most:2023sft}).
As a clear example, we plot a time series of three-dimensional renderings of magnetic field lines overplotted on two-dimensional slices of entropy per unit mass, $s = k_{\rm B}/[m_{\rm p} (\gamma-1)] \log(P/\rho^\gamma)$, for the tilted simulation in Figure \ref{fig:3D_fieldlines}.
Here $k_{\rm B}$ is Boltzmann's constant and $m_{\rm p}$ is the proton mass.
In this time series, not all magnetic field lines are shown, but only those which pass through at least one black hole event horizon \emph{and} which pass within $10 \rg$ of both black holes.
Field lines that pass through only the first black hole are colored yellow, field lines that pass through only the second black hole are colored blue, and field lines that pass through both are colored green.
In the particular instance shown in Figure \ref{fig:3D_fieldlines}, field lines anchored to the first black hole initially pass close to the second black hole.
These field lines start out in a relatively close bundle with predominantly straight field lines (Panel 1). 
The second black hole then captures several of these field lines (Panel 2), after which they get significantly twisted and inflate (Panels 3--4).
This twisting is caused by a relative rotation rate in the co-orbiting frame.\footnote{For each black hole the relative twisting motion depends on $\Omega_{\rm BH}-\Omega_{\rm orbit}$, where $\Omega_{\rm BH}$ $(\Omega_{\rm orbit})$ is the effective rotation rate of the black hole (orbit). As a consequence, twisting can even happen for irrotational black holes ($\Omega_{\rm BH}=0$).} 
This process ultimately leads to reconnection and an eruption that can potentially power a flare (see also \citealt{Yuan:2019mdb,Most:2020ami}).
When this happens, a large amount of magnetic energy is released, heating up the nearby gas to mildly relativistic temperatures [$\Theta_{\rm p} \equiv k_{\rm B} T_{\rm g} /(m_{\rm p} c^2) \lesssim 1$, where $T_{\rm g}$ is the gas temperature] and propelling it out to larger radii (Panels 5--6).
After this event the number of field lines connecting the two black holes is reduced (Panel 6) and ultimately goes back to $\sim 0$.  
In particular, the launching of the eruption is associated with the transient formation of a trailing current sheet; in other contexts the reconnection of such trailing current sheets has been shown to power subsequent high-energy emission (e.g., \citealt{Parfrey:2014wga,Most:2020ami}).
We isolate the current sheet for this particular eruption in Appendix \ref{app:recon_bridge} and show that it has negligible guide field (that is, all three components of the magnetic field change sign across the sheet) and is characterized by an upstream $\sigma$ of $\sim $ a few.

These types of eruptions in binary black hole systems differ from those containing at least one neutron star in an important way:
closed magnetic flux tubes threading black holes decay or open up exponentially with time and are thus short lived \citep{MacDonald:1982zz,Lyutikov2011,Bransgrove2021}. As a consequence, black holes fundamentally require accretion of plasma to both form and sustain magnetic loops threading their horizons \citep{Gralla2014}. 
This means that black hole binary eruptions are directly affected by the turbulent gas flow, and as a result are more stochastic in nature compared with neutron star-black hole or neutron-star neutron-star eruptions where there is a natural periodicity to the formation of connected field lines (with eruptions happening $\sim$ twice per orbit, \citealt{Cherkis2021,Most2022,Most2023}). 
How often flaring events occur in the black hole binary case depends crucially on the ratio between the time-averaged coherence length of the magnetic field in the orbital plane and the binary separation distance.
When this ratio is small (i.e., larger separation distances or earlier times), magnetic bridges can only form between the black holes when turbulence stochastically generates an instantaneously coherent field.
In this regime flaring events are rare.
When the ratio is closer to unity or larger (i.e., at shorter separation distances or later times closer to merger), magnetic bridges form constantly, and flaring events/eruptions are more frequent.  
In practice, this means that there will be some transition time when eruptions go from being rare to frequent, and possibly from stochastic to quasi-periodic as the binary separation distance decreases.  
In our simulations, this transition happens relatively close to merger, at separation distances $\lesssim$ 10--15$\rg$.  

We caution against over-generalizing this result on eruption recurrence rates without further study over a wider range of parameter space.  
The coherence length of the magnetic field in the orbital plane likely depends on the assumed magnetic field topology and assumed angular momentum of the gas (larger angular momentum can coherently wrap the magnetic field in the toroidal direction, e.g., in the circumbinary disk scenario).
Since we initialized the system with zero angular momentum and magnetic field purely in the vertical direction, the later quasi-periodic phase associated with frequency eruptions may not begin until particularly late times in our simulations.
This is in contrast to circumbinary disk scenarios where coherent loops could be accreted more often at larger separations (similar to \citealt{Parfrey:2014wga}). 

We find that when the black holes have strong jets (e.g., at most times in the fiducial simulation), magnetic bridges are prevented from forming and these types of eruptions do not occur.
This is because any field line that is accreted onto one black hole is generally twisted up and assimilated onto the same black hole's jet instead of lingering long enough to be accreted by the other black hole.
Even when such field lines \emph{do} get relatively close to the other black hole they tend to be blown away by that black hole's jet instead of being accreted. 
Thus, these magnetic bridge eruptions happen only when the jet is suppressed, whether that be from a lower magnetic flux supply (as in the tilted and $r_{\rm BH}=500\rg$ simulations), from low or no black hole spin (as in the $\chi=0$ simulation), or by the late-time suppression of $\phi_{\rm BH}$ caused by accelerating orbital velocities (as in the fiducial simulations).

Quantifying the electromagnetic emission associated with the eruptions discussed in this subsection is difficult without a fully general relativistic radiative transfer calculation on top of a model for the plasma magnetization, density, and content (i.e., whether it is electron-positron pairs or electron-ion pairs) as well as the magnetic field strength near the event horizon.  
This is because quantities such as the Poynting flux or energy outflow measured from the simulations are not necessarily a good proxy for radiative signatures.
Moreover, even if the total energy released is small compared to the overall luminosity of the system, it is likely that the hot plumes produced by eruptions radiate at much higher frequencies than the bulk of the accretion flow and would thus dominate the emission in that regime.
Likewise, the strongest electromagnetic signature may come from accelerated, nonthermal particles.  
These particles are fundamentally not captured in our ideal GRMHD fluid approximation and thus understanding their acceleration and emission in detail will require more localized particle-in-cell simulations.

With those caveats in mind, we still find it instructive to make some order-of-magnitude estimates.  
In Appendix \ref{app:recon_bridge} we roughly estimate the energetics of these events, finding that the total energy released by reconnection is 
 \begin{equation}
 \begin{aligned}
    \frac{\left( L_{\rm rec}\right)_{\rm bridge}}{L} \approx\ & 0.08\  \left(\left. \frac{u_{b,\rm rec} r_{\rm g}^2}{|\dot M |c} \right/ 0.075 \right)_{r=r_{\rm H}} \\
    &\times\ \left(\frac{r_{\rm H}}{0.67 \rg}\right)^2 \left[ \frac{\log\left(l_{\rm rec}/r_{\rm H}\right)}{2.7}\right],
    \end{aligned}
\end{equation} 
or 
 \begin{equation}
 \begin{aligned}
    \left( L_{\rm rec}\right)_{\rm bridge} \approx\ & 10^{43}\ \frac{\mathrm{erg}}{\mathrm{s}}\ \left(\left. \frac{u_{b,\rm rec} r_{\rm g}^2}{|\dot M |c} \right/ 0.075 \right)_{r=r_{\rm H}} \left(\frac{r_{\rm H}}{0.67 \rg}\right)^2 \\
    &\times\  \left[ \frac{\log\left(l_{\rm rec}/r_{\rm H}\right)}{2.7}\right] f_{\rm Edd}\left(\frac{M}{10^6 M_\odot}\right) \left(\frac{\eta_{\rm rad}}{0.1}\right)^{-1},
\end{aligned}
\end{equation}
emitted as synchrotron radiation at a frequency of 
\begin{equation}
\begin{aligned}
   (\nu_{\rm synch})_{\rm bridge} \approx\ & 3 \times 10^{20}\ \mathrm{Hz}\ \left(\frac{\sigma_{\rm bridge}}{4}\right)^2 \left( \left. \frac{u_{b, \rm rec} \rg^2}{|\dot M| c} \right/ 0.075 \right)^{1/2}\\ 
   &\times\ f_{\rm Edd}^{1/2}\ \left(\frac{M}{10^6 M_\odot}\right)^{-1/2} \left(\frac{\eta_{\rm rad}}{0.1}\right)^{-1/2}.
\end{aligned}
\end{equation}
where $r_{\rm H}$ is the event horizon radius of one of the black holes.
This frequency is $\gtrsim 10^3$ times the thermal synchrotron peak that we estimate in Appendix \ref{app:recon_therm} but $\sim$ $10^{2\textrm{--}8}$ times less than the characteristic synchrotron frequency of the jet-jet emission that we estimate in Appendix \ref{app:recon_jet} depending on $M$ and $f_{\rm Edd}$, where larger values of $(\nu_{\rm synch})_{\rm jet}/(\nu_{\rm bridge})_{\rm jet} $ correspond to smaller Eddington ratios and/or larger black hole masses.
That means that the three emission processes (one thermal and the others nonthermal) are likely distinguishable from each other and could be independently detectable.

\section{Discussion and Conclusions}
\label{sec:disc_conc}
We have presented a suite of four 3D GRMHD simulations of SMBH binary accretion in low angular momentum environments.  
We varied the black hole spin directions and magnitudes as well as the initial Bondi radius of the gas. 
The simulations were run up until just before merger, $\approx$ 4--6 $\times$ $10^4 M$ or $70-100$ orbits for initial separations of $25$--$27$ $\rg$. 
These are the longest run and largest separation distances studied to date in GRMHD simulations of binary black holes that include the event horizons.
While a large fraction of parameter space in SMBH binary accretion still remains unexplored by both theory and simulations, in this work we have studied a very limited range of that parameter space in order to highlight several key areas that could be fruitful for future study and the interpretation and prediction of observations.

We find that feedback from the black holes can partially inhibit the accumulation of net magnetic flux on the event horizon in certain parameter regimes (Figure \ref{fig:time_plots}).  
This is true even though the initial magnetic field geometry contains a significant amount of ordered vertical flux in all simulations and the resulting near-horizon flow is highly magnetized ($\beta \lesssim$ a few on average).
In particular, we have shown that simulations with black hole spins moderately tilted with respect to the orbital angular momentum and/or with flows that have an initially large Bondi radius reach saturated states with relatively lower horizon-penetrating dimensionless magnetic flux (compared with simulations using smaller Bondi radii and aligned spins).
This is because feedback through either misaligned jets or flux tube ejections is significant enough to regulate the amount of magnetic flux that ultimately reaches the black hole event horizons (Figure \ref{fig:corotating_slice}).
As a result, only one of the three simulations with nonzero black hole spin displays persistent and structured jets that reach large radii (Figure \ref{fig:3D_jet}).
The other simulations show intermittent, weak, and quickly dissipated jets that only occasionally reach larger radii.
Additionally, even the jets in the fiducial simulation do not reach $>100\%$ efficiencies like those typically seen in magnetically arrested flows, possibly due to enhanced dissipation/feedback associated with longer current sheets that are extended by the orbital motion of the black holes.

We have demonstrated two potential emission mechanisms that are unique to merging SMBH binary AGN (compared with single SMBH AGN).
Both involve magnetic reconnection and are likely to accelerate electrons to high energies and produce high-frequency electromagnetic emission with characteristic luminosities $\lesssim$ 10\% of the total thermal luminosity of the accretion flow.  
The first mechanism occurs when the black holes power persistent jets and the spins are aligned.
In this case, the two jets form an extended reconnection layer that dissipates magnetic energy and causes the jets to merge (Figure \ref{fig:jet_interaction}, Appendix \ref{app:recon_jet}).  
This is analogous to flux tube collisions in the solar corona (i.e., collisions of two footpoints at the solar surface), a probable cause of certain solar flares \citep{Linton2001}.
This reconnection layer is characterized by a highly magnetized upstream flow ($\sigma\gg 1$ and $\beta \ll 1$) with a moderately strong guide field of comparable strength to the reconnecting field.
Reconnection in this regime is known to be a source of high-energy particles that would radiate at higher frequencies than the accretion disk (as we estimate in Appendix \ref{app:recon}).

When the black holes have jets that are either weak or nonexistent, we have proposed and demonstrated a novel flaring mechanism. 
Analogous to coronal mass ejections in the sun or similar phenomena proposed in neutron star-neutron star and neutron star-black hole mergers \citep{Most:2020ami,Most2023}, ``magnetic bridge'' (connected flux tube) eruption events occur when magnetic field lines get accreted by both black holes and subsequently get twisted and torqued by the orbital motion, forming an equatorial current sheet (Figure \ref{fig:3D_fieldlines}, Appendix \ref{app:recon_bridge}).
When the tension on the field lines reaches a critical point, they break off through reconnection and create an unbound outflow of hot plasma. 
Unlike in mergers containing a neutron star, these flares require accreting plasma and thus are stochastic in nature, with occurrence rates that depend on the separation distance of the binary and the time-averaged coherence length of the magnetic field in the orbital plane.  
For large separation distances (short coherence lengths), flares are stochastic and more rare, relying on spontaneous, turbulent generation of coherent magnetic field.  
For short separation distances (long coherence lengths), eruptions are frequent and possibly even quasi-periodic.
In either regime, this flaring mechanism is a promising candidate for providing energetic observational signatures of SMBH binary in-spirals and mergers that are counterparts to merging low frequency gravitational wave sources.

\begin{acknowledgments}
We thank the anonymous referee for useful comments on the manuscript.
We thank A.A.\ Philippov, E.M.\ Guti\'errez,  and J.S.\ Heyl for useful discussions.
We acknowledge the support of the Natural Sciences and Engineering Research Council of Canada (NSERC), [funding reference number 568580]
Cette recherche a \'et\'e financ\'ee par le Conseil de recherches en sciences naturelles et en g\'enie du Canada (CRSNG), [num\'ero de r\'ef\'erence 568580].
L.C.\ is supported in part by
Perimeter Institute for Theoretical Physics.
Research at Perimeter Institute is supported in part by the Government of Canada through the Department of Innovation, Science and Economic Development Canada and by the Province of Ontario through the Ministry of Colleges and Universities. ERM acknowledges partial support by the National Science Foundation under grants No. PHY-2309210 and AST-2307394.
B.R.\ is supported by the Natural Sciences \& Engineering Research Council of Canada (NSERC), the Canadian Space Agency (23JWGO2A01), and by a grant from the Simons Foundation (MP-SCMPS-00001470). B.R. acknowledges a guest researcher position at the Flatiron Institute, supported by the Simons Foundation.
E.R.M.\ gratefully acknowledges the hospitality of the
Aspen Center for Physics, which is supported by National Science Foundation
grant PHY-2210452.

The computational resources and services used in this work were partially provided by facilities supported by the VSC (Flemish Supercomputer Center), funded by the Research Foundation Flanders (FWO) and the Flemish Government – department EWI and by Compute Ontario and the Digital Research Alliance of Canada (alliancecan.ca).  
\end{acknowledgments}

\software{{\tt Athena++} \citep{White2016,Athenapp},
{\tt CBwaves} \citep{cbwaves},
{\tt matplotlib} \citep{Hunter:2007},}

\bibliographystyle{aasjournal}
\bibliography{mad_binaries}

\appendix

\section{On The Use of the Term ``Magnetically Arrested''}
\label{app:MAD}
The term ``magnetically arrested'' has been widely adopted to describe a class of simulations in GRMHD \citep{Narayan2003,Igumenshchev2003,Narayan2012,Sasha2011,EHT5,EHT_SGRA_5,Chatterjee2022}.  
Generally, these simulations are characterized by several key features:
\begin{itemize}
\item The accretion of a significant amount of poloidal magnetic flux onto the event horizon that tends to form a transient, reconnecting, equatorial current sheet.
\item The saturation of dimensionless magnetic flux, $\phi_{\rm BH} = \Phi_{\rm BH}/\sqrt{|\dot M|} $ at values between 40--60, with cycles of slow increase followed by rapid dissipation.
\item Jets with efficiencies $\gtrsim$ 100 \% relative to the accretion power, $|\dot M| c^2$, when the black hole is rapidly spinning.
\item The frequent ejection of low density, highly magnetized ``flux tubes'' by the reconnecting midplane current sheet that regulate the amount of accreted magnetic flux.
\item A highly magnetized inner accretion flow, with accretion in the innermost region proceeding in thin streams.
\item Accretion disks with sub-Keplerian orbital velocities.
\end{itemize}

In previous literature, GRMHD simulations that start with an initial torus of gas (e.g., \citealt{Fishbone1976,Chakrabarti1985,Penna2013}), displayed a clear dichotomy between SANE and MAD flows; simulations either showed all of the above listed properties or none.  
Recent work on more spherical distributions of gas studying a larger dynamical range of accretion \citep{Ressler2021,Lalakos2022,Kwan2023,Lalakos2024,Galishnikova2024}, however, has complicated this picture.
These authors have found that in certain parameter regimes you can have most of the traditional MAD features but with moderately lower horizon-penetrating magnetic fluxes ($\phi_{\rm BH} \sim$ 10--30) and without particularly strong jets.
This is not due to a lack of available magnetic flux supply (as is the case for SANE flows) but due to a change in magnetic flux transport caused by feedback from either jets or flux tube ejections.  
Stronger feedback is found for simulations with larger Bondi radii, which results in less net magnetic flux reaching the event horizons and weaker jets.
We would argue that such simulations are qualitatively in the same accretion state as those with larger horizon-penetrating magnetic fluxes and stronger jets.
In both cases, the magnetic flux threading the event horizon reaches a clear saturation where inflow and outflow of magnetic flux are roughly balanced in a time-averaged sense.
Moreover, this saturation is governed by the same physical mechanisms.
Inflow proceeds by accretion and outflow proceeds by the ejection of flux tubes and/or by jets (i.e., feedback).\footnote{Note that since the feedback can also affect the inflow of magnetic fields, this is a nonlinear process.}
In contrast, the accretion of magnetic flux in SANE flows is generally not regulated by this feedback (since there are generally no equatorial current sheets that drive flux tube ejection), and $\phi_{\rm BH}$ does not saturate at a mean value but instead shows more secular variability.

We therefore propose that instead of being a distinct state with universal characteristics, magnetically arrested accretion flows fall on a spectrum.
When flux tube/jet feedback is strong, $\phi_{\rm BH}$ can drop to moderate values and the jets are weak, but when feedback is weak $\phi_{\rm BH}$ can reach maximal values and the jets can be $> 100\%$ efficient.
In this sense, we could describe all of the flows in our simulations as ``magnetically arrested,'' as they all display reconnection-driven flux tube ejection and highly magnetized inner accretion flows. 
If we use the more traditional and strict definition of the MAD state, however, none of our simulations are fully MAD as even the most powerful jets are well below 100\% efficiency.

Because of this ambiguity, in the main text we do not definitively refer to any of our simulations as either MAD or SANE.  
Instead we focus on highlighting the qualitative and quantitative features of MAD flows that are or are not present in each.

\section{Measuring The Properties of Magnetic Reconnection and Estimating Electromagnetic Emission}
\label{app:recon}
\begin{figure}
\includegraphics[width=0.45\textwidth]{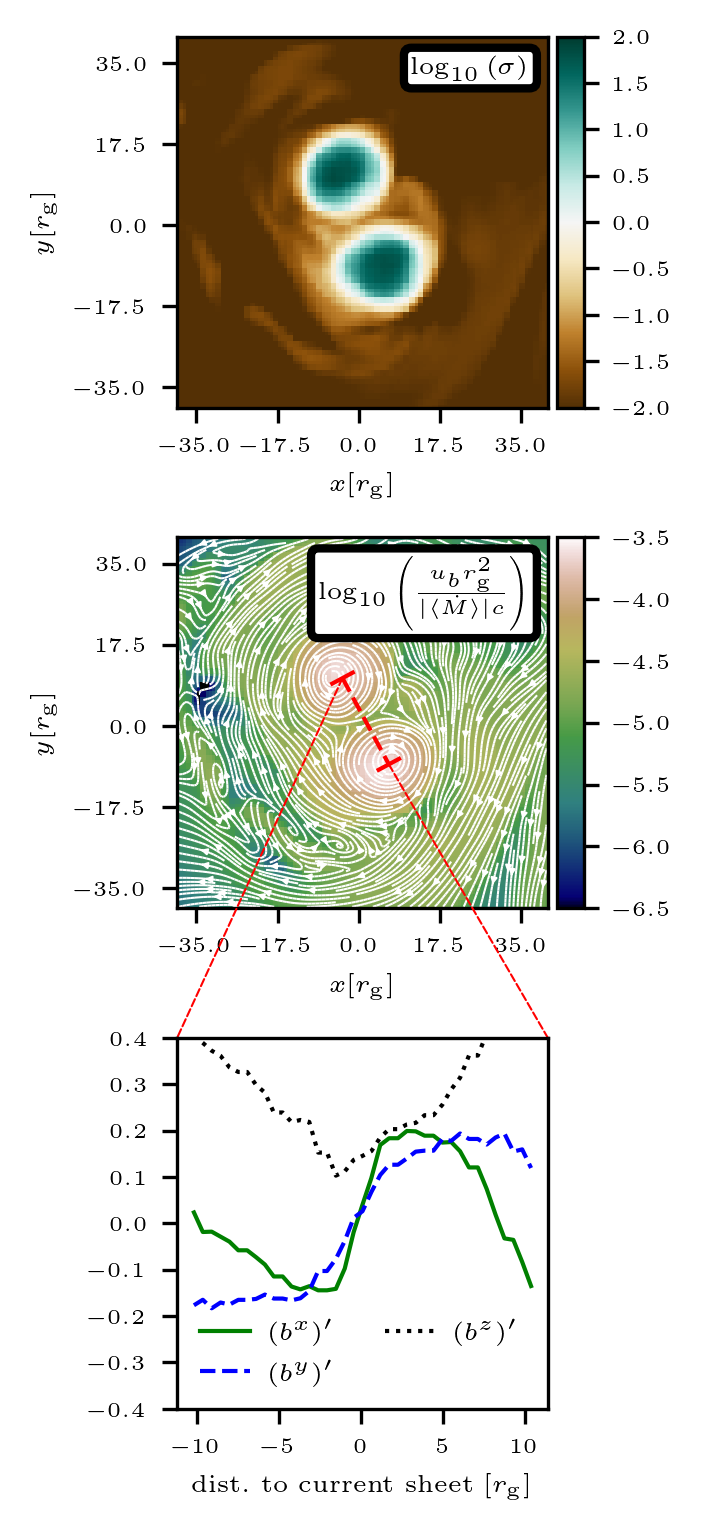}
\caption{Properties of the magnetic reconnection layer formed by the two interacting jets at the time shown in Figure \ref{fig:jet_interaction}.
Top: Two-dimensional slice of $\sigma$ through the $z=50 \rg$ plane.
Middle: Same as the top panel but for magnetic energy density normalized to the time-averaged accretion rate, $u_b \rg^2/(|\langle \dot M \rangle| c)$ with magnetic field lines overplotted.
Bottom: One-dimensional profiles through the reconnection layer (along the line indicated in the middle panel) of the three spatial components of the co-moving magnetic field boosted into the jet frame, $\left(b^\mu\right)^\prime$.
The upstream flow is characterized by $\sigma \approx 100$, the imposed numerical limit (meaning that in reality it would be much larger) and a guide field of comparable strength to the reconnecting field.
}
\label{fig:current_sheet_jet}
\end{figure}

We can roughly estimate the energy released by reconnection using $L_{\rm rec} \approx u_{b,\rm rec} l_{\rm rec}^2 \beta_{\rm rec} c $, where $u_{b,\rm rec} = b_{\rm rec}^2/2$ is the magnetic energy density of the reconnecting magnetic field (in Lorentz-Heaviside units), $l_{\rm rec}$ is the length of the reconnection layer (not to be confused with the width or thickness, which in reality is microscopic), and $\beta_{\rm rec}$ is the reconnection rate.
Particle-in-cell simulations generally find $\beta_{\rm rec} \approx 0.1$ \citep{SironiSpitkovsky2014,2014PhRvL.113o5005G,2016ApJ...816L...8W}, so if we parameterize the overall luminosity of the accretion flow with a radiative efficiency of $\eta_{\rm rad}$ as $L = \eta_{\rm rad}\dot M c^2$, we can write:
\begin{equation}
\label{eq:Lrec}
\frac{L_{\rm rec}}{L} \approx\   \left(\frac{u_{b,\rm rec} r_{\rm g}^2}{|\dot M |c}\right) \left(\frac{\eta_{\rm rad}}{0.1}\right)^{-1} \left(\frac{l_{\rm rec}}{r_{\rm g}}\right)^2.
\end{equation}
Here $u_{b,\rm rec} r_{\rm g}^2/(|\dot M| c)$ is essentially a dimensionless and scale-free measure of the magnetic field strength relative to the accretion power.  
Both $u_{b,\rm rec} r_{\rm g}^2/(|\dot M| c)$ and $l_{\rm rec}$ we can directly measure from the simulations for each reconnection scenario. 

To obtain estimates in cgs units, we can parameterize the Luminosity as $L = f_{\rm Edd} L_{\rm Edd}$, where $f_{\rm Edd}$ is the Eddington ratio and $L_{\rm Edd}$ is the Eddington luminosity.  
This results in 
\begin{equation}
\label{eq:Lrec_cgs}
L_{\rm rec} \approx\  1.3 \times 10^{44}\  \frac{\mathrm{erg}}{\mathrm{s}}\ \left(\frac{u_{b,\rm rec} r_{\rm g}^2}{|\dot M |c}\right) \left(\frac{l_{\rm rec}}{r_{\rm g}}\right)^2 f_{\rm Edd}\left(\frac{M}{10^6 M_\odot} \right)\left(\frac{\eta_{\rm rad}}{0.1}\right)^{-1} .
\end{equation}
Furthermore, we can parameterize the magnetic field strength in a similar way:
\begin{equation}
    B_{\rm cgs} \approx\ 7 \times 10^6\  \mathrm{G}\  \left( \frac{u_{\rm b} \rg^2}{|\dot M| c} \right)^{1/2} f_{\rm Edd}^{1/2} \left(\frac{M}{10^6 M_\odot}\right)^{-1/2} \left(\frac{\eta_{\rm rad}}{0.1}\right)^{-1/2}  
\end{equation}
corresponding to an electron gyro-frequency $[\omega_{\rm b} = \mathrm{e}B/(m_{\rm e}c)]$:
\begin{equation}
\omega_{\rm b} \approx\ 1.2 \times 10^{14}\ \frac{\mathrm{rad}}{\mathrm{s}}\ \left( \frac{u_{\rm b} \rg^2}{|\dot M| c} \right)^{1/2} f_{\rm Edd}^{1/2} \left(\frac{M}{10^6 M_\odot}\right)^{-1/2} \left(\frac{\eta_{\rm rad}}{0.1}\right)^{-1/2}.
\end{equation}

\subsection{Interacting Jets}
\label{app:recon_jet}

In Figure \ref{fig:current_sheet_jet} we isolate the properties of the jet-jet reconnection layer at $z=50 \rg$ by plotting two-dimensional slices of $\sigma$ and $u_{b,\rm rec} r_{\rm g}^2/(|\dot M| c)$ as well as a one-dimensional profile of the three spatial components of the co-moving magnetic field boosted to the jet frame.  
The boost is performed using the average coordinate velocity of the jet at this distance, 0.7$c$ in the $+z$ direction.
We find that the upstream $u_{b,\rm rec} r_{\rm g}^2/(|\dot M| c)$ is $\sim$ $5 \times 10^{-5}$ while $l_{\rm rec}$ is $\sim$ $20\rg$.
This implies that 
\begin{equation}
        \frac{\left( L_{\rm rec}\right)_{\rm jet}}{L} \approx\ 0.02\   \left(\left. \frac{u_{b,\rm rec} r_{\rm g}^2}{|\dot M |c} \right/ 5 \times 10^{-5}\right) \left(\frac{l_{\rm rec}}{ 20 \rg}\right)^2,
\end{equation}
or 
\begin{equation}
\begin{aligned}
        \left( L_{\rm rec}\right)_{\rm jet} \approx\ & 3\times 10^{42}\ \frac{\mathrm{erg}}{\mathrm{s}}\  \left(\left. \frac{u_{b,\rm rec} r_{\rm g}^2}{|\dot M |c} \right/ 5 \times 10^{-5}\right) \left(\frac{l_{\rm rec}}{ 20 \rg}\right)^2 \\
        &\times f_{\rm Edd}\left(\frac{M}{10^6 M_\odot}\right) \left(\frac{\eta_{\rm rad}}{0.1}\right)^{-1}.
\end{aligned}
\end{equation}

The upstream $\sigma$ is the ceiling value set by the density floor, meaning that in reality it depends on how the jet is mass-loaded. 
We can estimate a more realistic value by using the \citet{Goldreich1Julian} number density, $n_{\rm GJ}$, needed to sustain the induced electric field in the jet (see also discussion in Section 4.1 of \citealt{Ripperda2022}), $\sigma_{e,\rm GJ} = b^2/(m_{\rm e}n_{\rm GJ} c^2)$, where $n_{\rm GJ} = \lambda \Omega_{\rm BH} b  /(2 \mathrm{\pi} c e)$, $\lambda$ is a multiplicity factor $\lesssim 10^{3}$ \citep{Mosci2011,Chen2020,Crinquand2020}, and $\Omega_{\rm BH} = \chi c /(2r_{\rm H})$.  
This results in
\begin{equation}
\label{eq:sige_gs}
\begin{aligned}
    \left(\sigma_{\rm e}\right)_{\rm jet} \approx\ & 2 \times 10^{10}\ \left( \left. \frac{u_{b,\rm rec} \rg^2}{|\dot M| c} \right/ 5 \times 10^{-5}\right)^{1/2} \left(\frac{\lambda}{10^3}\right)^{-1} 
    \\ & \times f_{\rm Edd}^{1/2} \left(\frac{M}{10^6 M_\odot}\right)^{1/2} \left(\frac{\eta_{\rm rad}}{0.1}\right)^{-1/2},  
    \end{aligned}
\end{equation}
where we have substituted $u_{b,\rm rec} r_{\rm g}^2/(|\dot M| c)$ $\sim$ $5 \times 10^{-5}$ and the approximate upper limit on $\lambda$ of $10^3$.  
Without radiative cooling, electrons accelerated by reconnection would reach $\Gamma_{\rm e} \approx \sigma_{\rm e}$.  
However, radiative back reaction on the particles will limit their Lorentz factors to be less than a critical value, $\Gamma_{\rm e,cool}$, obtained by equating the radiative drag force of a particle with the force provided by the accelerating electric field, $\Gamma_{\rm e,cool}^2 \approx 3 m_{\rm e}^2 c^4/(2e^3 \sqrt{b^2})$ (e.g., \citealt{Uzdensky2011,Ripperda2022}).
For our parameters we find
\begin{equation}
\label{eq:gamma_cool}
\begin{aligned}
\Gamma_{\rm e,cool} \approx\ & 4 \times 10^{5}\ \left( \left. \frac{u_{b, \rm rec} \rg^2}{|\dot M| c} \right/ 5\times  10^{-5}\right)^{-1/4} \\ 
& \times f_{\rm Edd}^{-1/4} \left(\frac{M}{10^6 M_\odot}\right)^{1/4} \left(\frac{\eta_{\rm rad}}{0.1}\right)^{1/4}.
\end{aligned}
\end{equation}
For all Eddington ratios $\gtrsim 10^{-7}$ and $\lambda\lesssim 10^3$, we estimate that $\Gamma_{\rm e,cool} < \sigma_{\rm e}$, meaning that particle acceleration in the jet-jet reconnection layer is radiatively limited.
As a result we can estimate the characteristic photon frequency at which the accelerated electrons would emit synchrotron radiation using $\nu_{\rm synch} = \Gamma_{\rm e,cool}^2 \omega_{\rm b}/(2 \mathrm{\pi})$:
\begin{equation}
\label{eq:nusynch_jet}
(\nu_{\rm synch})_{\rm jet} \approx\ 2.5 \times 10^{22}\mathrm{Hz}.
\end{equation}
This is the so-called ``synchrotron burn-off limit'' and is independent of any other parameters.

In the bottom panel panel of Figure \ref{fig:current_sheet_jet}, we find that the jet-jet reconnection occurs with significant guide field, $\sqrt{b_z b^z}/\sqrt{b_x b^x + b_y b^y} \sim 1$.  
As discussed in the main text, this level of guide field will moderately limit the maximum Lorentz factor of the accelerated electrons, though precisely quantifying this effect in the regime of interest is still an ongoing area of research.

\begin{figure*}
\includegraphics[width=0.95\textwidth]{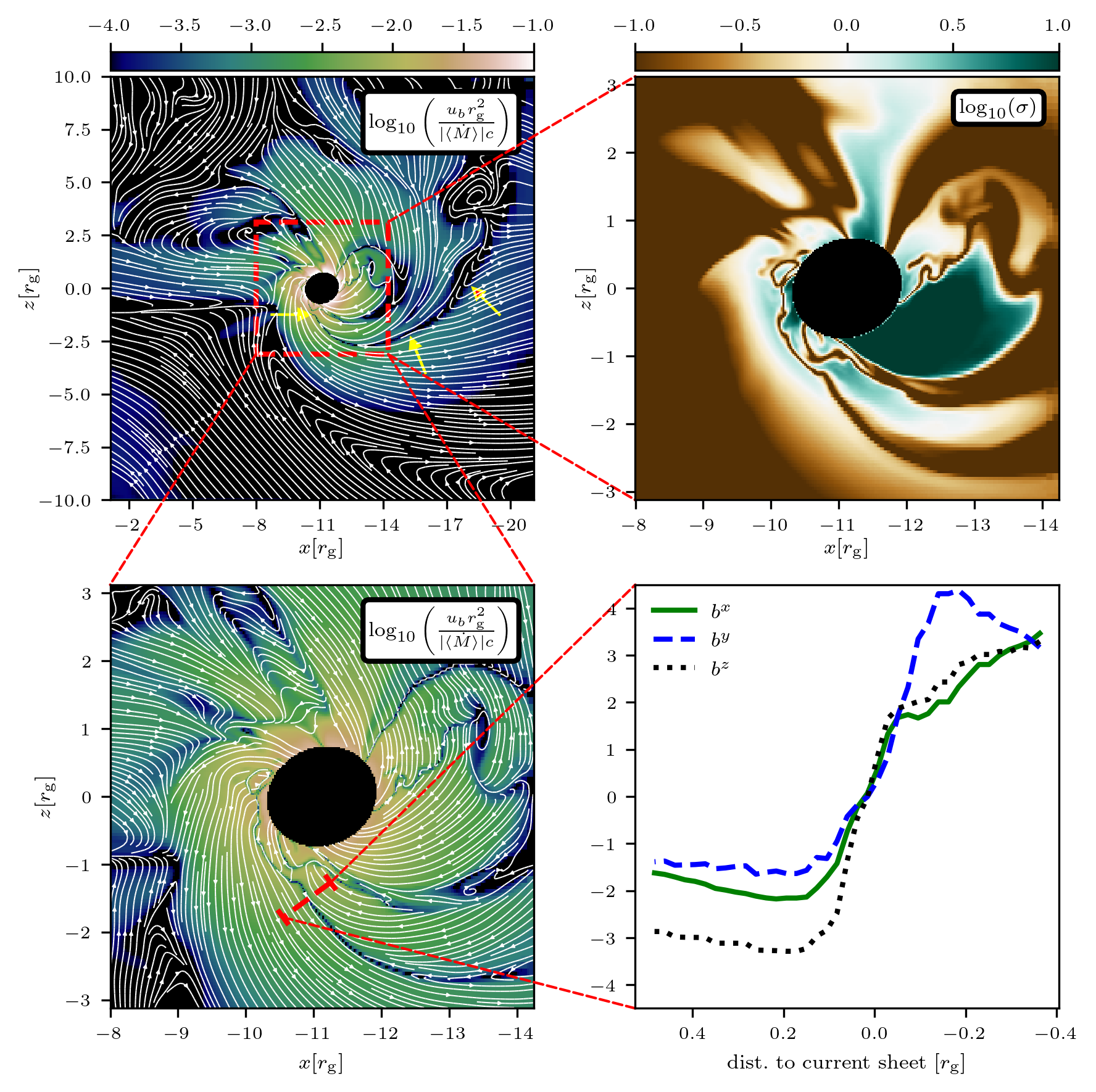}
\caption{Properties of the magnetic reconnection layer formed just before the eruption event in the tilted simulation plotted in Figure \ref{fig:3D_fieldlines}.
Top left: Two-dimensional slice centered on the second black hole of magnetic energy density normalized to the time-averaged accretion rate, $u_b \rg^2/(|\langle \dot M \rangle| c)$, with the reconnection layer indicated by yellow arrows.
Top right: Two-dimensional slice of $\sigma=b^2/\rho$.
Bottom left: same as top left but on a $\sim$ 4 times smaller scale.
Bottom right: 1D profiles through the reconnection layer (along the line indicated in the bottom left panel) of the three spatial components of the co-moving magnetic field, $b^\mu$.
The available energy for reconnection is highest near the horizon, with $u_b \rg^2/(|\langle \dot M \rangle| c)$ $\sim$ 0.075, where the layer is surrounding by plasma with $\sigma \sim $ a few.
Furthermore, there is negligible guide field, with all three spatial components of $b^\mu$ changing sign across the layer.
}
\label{fig:current_sheet_bridge}
\end{figure*}

\subsection{Magnetic Bridge Eruptions}
\label{app:recon_bridge}

Isolating the current sheet associated with the eruption shown in Figure \ref{fig:3D_fieldlines}
 of the main text is more difficult than the jet interaction layer because it doesn't align with any particular coordinate plane.  
 Part of it can be seen, however, in the $x-z$ plane centered on the second black hole as we show in Figure \ref{fig:current_sheet_bridge}.
 In this figure we plot two-dimensional slices of $u_{b,\rm rec} r_{\rm g}^2/(|\dot M| c)$ on two different scales, a two-dimensional slice of $\sigma$ on the smaller scale, and a one-dimensional profile of the three spatial components of the co-moving magnetic field.
 The current sheet is seen as a region of low $b^2$ that attaches to the black hole near the midplane on the left side of the figure and then loops around the bottom of the black hole and turns towards positive $z$.  
 We have confirmed that this is the current sheet associated with the flux tube plotted in Figure \ref{fig:3D_fieldlines}.
 Since the magnetic field strength in this flux tube generally decays with distance from the black hole as $b^2\ \tilde \propto\ l^{-2}$, the dissipated power from reconnection is dominated by the near-horizon region.  
 More precisely, we should differentiate Equation \eqref{eq:Lrec} with respect to $l_{\rm rec}$ and integrate along the length of the current sheet, giving
 \begin{equation}
 \label{eq:Lrec_bridge}
     L_{\rm rec}/L \approx\ \left(\frac{u_{b,\rm rec} r_{\rm g}^2}{|\dot M |c}\right)_{r=r_{\rm H}} \left(\frac{r_{\rm H}}{r_{\rm g}}\right)_{r=r_{\rm H}}^2 2 \log\left(\frac{l}{r_{\rm H}}\right) \left(\frac{\eta_{\rm rad}}{0.1}\right)^{-1}.
 \end{equation}
 From Figure \ref{fig:current_sheet_bridge} we can estimate $u_{b,\rm rec} r_{\rm g}^2/(|\dot M| c)$ at the event horizon as $\approx$ $0.075$ and $l_{\rm rec} \approx 10 \rg$. 
 Since the event horizon radius for this simulation is $r_{\rm H} \approx 0.67 \rg$, we estimate 
 \begin{equation}
  \begin{aligned}
    \frac{\left( L_{\rm rec}\right)_{\rm bridge}}{L} \approx\  & 0.08  \left(\left. \frac{u_{b,\rm rec} r_{\rm g}^2}{|\dot M |c} \right/ 0.075 \right)_{r=r_{\rm H}} \\
    &\times\ \left(\frac{r_{\rm H}}{0.67 \rg}\right)^2 \left[ \frac{\log\left(l_{\rm rec}/r_{\rm H}\right)}{2.7}\right],
    \end{aligned}
\end{equation} 
 or 
 \begin{equation}
 \begin{aligned}
    \left( L_{\rm rec}\right)_{\rm bridge} \approx\ & 10^{43}\ \frac{\mathrm{erg}}{\mathrm{s}}\ \left(\left. \frac{u_{b,\rm rec} r_{\rm g}^2}{|\dot M |c} \right/ 0.075 \right)_{r=r_{\rm H}} \left(\frac{r_{\rm H}}{0.67 \rg}\right)^2 \\
    &\times\  \left[ \frac{\log\left(l_{\rm rec}/r_{\rm H}\right)}{2.7}\right] f_{\rm Edd}\left(\frac{M}{10^6 M_\odot}\right) \left(\frac{\eta_{\rm rad}}{0.1}\right)^{-1}.
    \end{aligned}
 \end{equation}

 From Figure \ref{fig:current_sheet_bridge} it is also clear that the upstream near-horizon $\sigma$ is $\sim$ a few and that there is negligible guide field (that is, all three spatial components of the field change sign across the current sheet).
 In this regime we expect electrons to be accelerated to Lorentz factors of $\Gamma_{\rm e}$ $\sim$ $\sigma_{\rm e}$, where $\sigma_{\rm e} \approx \sigma m_{\rm p}/m_{\rm e}$, $m_{\rm e}$ and $m_{\rm p}$ are the electron and proton masses, respectively, and we have assumed that the plasma is composed of hydrogen ions.
 Using $\sigma \approx 4$, we find $\Gamma_{\rm e}$ $\sim$ $7300 (\sigma_{\rm bridge}/4$).
For all reasonable black hole masses and Eddington ratios, this Lorentz factor is smaller than the synchrotron cooling limit (Equation \ref{eq:gamma_cool}). 
Thus we estimate:
\begin{equation}
\label{eq:nusynch_bridge}
\begin{aligned}
   (\nu_{\rm synch})_{\rm bridge} \approx\ & 3 \times 10^{20}\ \mathrm{Hz}\ \left(\frac{\sigma_{\rm bridge}}{4}\right)^2 \left( \left. \frac{u_{b, \rm rec} \rg^2}{|\dot M| c} \right/ 0.075 \right)^{1/2}\\ 
   &\times\ f_{\rm Edd}^{1/2}\ \left(\frac{M}{10^6 M_\odot}\right)^{-1/2} \left(\frac{\eta_{\rm rad}}{0.1}\right)^{-1/2}.
\end{aligned}
\end{equation}

\subsection{Thermal Emission}
\label{app:recon_therm}

We can also make a rough estimate of the thermal emission properties of the accretion flow.  
 If we assume that the thermal temperature of the electrons in the near-horizon flow is some fraction $f_{\rm e} \le 1$ of the ion temperature, then this can be compared to the ``thermal'' Lorentz factor of $\Theta_{\rm e} \equiv k_{\rm B} T_{\rm e}/(m_{\rm e} c^2) = (m_{\rm p}/m_{\rm e})f_{\rm e}P/(2\rho) $, again assuming hydrogen ions.  
 Measuring this temperature in a mass-weighted average in our simulations results in $\langle \Theta_{\rm e} \rangle_\rho \lesssim$ $200f_{\rm e}$.
 Note that this value should be taken as an upper-limit to the actual emitting temperature of the thermal electrons since gas at slightly larger radii likely contribute significantly.
This results in a characteristic synchrotron frequency of $\nu_{\rm synch} = \Theta_{\rm e}^2 \omega_{\rm b}/(2\mathrm{\pi})$:
\begin{equation}
\begin{aligned}
       (\nu_{\rm synch})_{\rm th} \approx\ & 2 \times 10^{17} \ \mathrm{Hz}\ \left(\frac{\Theta_{\rm e}}{200} \right)^2 \left( \left. \frac{u_{b} \rg^2}{|\dot M| c} \right/ 0.075 \right)^{1/2} \\
       & \times f_{\rm Edd}^{1/2} \left(\frac{M}{10^6 M_\odot}\right)^{-1/2} \left(\frac{\eta_{\rm rad}}{0.1}\right)^{-1/2},
\end{aligned}
\end{equation}
where we have used the near-horizon value of $u_b \rg^2/(|\dot M|c)\sim 0.075$ as in Appendix \ref{app:recon_bridge}.
 
 We can now compare this frequency to the characteristic synchrotron frequency of the reconnection events.
 For the jet-jet reconnection layer we estimate
 \begin{equation}
 \begin{aligned}
     \frac{(\nu_{\rm synch})_{\rm jet}}{(\nu_{\rm synch})_{\rm th}} \approx\ & 10^5\ \left(\frac{\Theta_{\rm e}}{200} \right)^{-2} 
  \left(\left. \frac{u_{b} \rg^2}{|\dot M| c} \right/ 0.075 \right)^{-1/2} \\
  & \times f_{\rm Edd}^{-1/2} \left(\frac{M}{10^6 M_\odot}\right)^{1/2} \left(\frac{\eta_{\rm rad}}{0.1}\right)^{1/2},
\end{aligned}
 \end{equation}
 while for the magnetic bridge eruptions we estimate 
  \begin{equation}
     \frac{(\nu_{\rm synch})_{\rm bridge}}{(\nu_{\rm synch})_{\rm th}} \approx\ 1.3  \times 10^3\  \left(\frac{\Theta_{\rm e}}{200} \right)^{-2} \left(\frac{\sigma_{\rm bridge}}{4}\right)^2.
 \end{equation}

 We can also compare $(\nu_{\rm synch})_{\rm jet}$ and $(\nu_{\rm synch})_{\rm bridge}$ directly, finding:
 \begin{equation}
     \begin{aligned}
              \frac{(\nu_{\rm synch})_{\rm jet}}{(\nu_{\rm synch})_{\rm bridge}} \approx\ & 80\ \left(\frac{\sigma_{\rm bridge}}{4}\right)^{-2}
  \left(\left. \frac{u_{b} \rg^2}{|\dot M| c} \right/ 0.075 \right)^{-1/2} \\
  & \times f_{\rm Edd}^{-1/2} \left(\frac{M}{10^6 M_\odot}\right)^{1/2} \left(\frac{\eta_{\rm rad}}{0.1}\right)^{1/2}.
     \end{aligned}
 \end{equation}

\subsection{Summary}
To summarize, we generally estimate that both jet-jet reconnection events and magnetic bridge reconnection events  in our simulations can produce a significant amount of luminosity, between a few to ten percent of the total luminosity provided by the bulk of the accretion flow.  
We predict that this luminosity would appear as synchrotron emission at higher frequencies than the thermal synchrotron peak by factors of $>$ $10^3$.  
For all reasonable values of $f_{\rm Edd} \gtrsim 10^{-7}$ and $10^6 M_\odot  \lesssim M \lesssim 10^{10} M_\odot$, our analysis finds that the jet-jet emission will appear at higher frequencies than the magnetic bridge eruption emission (by at least a factor of $\gtrsim 80$ and by as much as $\gtrsim 10^7$ depending on $M$ and $f_{\rm Edd}$), with the latter having a slightly higher luminosity (a factor of $\sim$ 2 for the particular parameters we measure in the simulations).
Additionally, inverse Compton scattering has the potential to up-scatter photons to even higher frequencies than those estimated here.

Finally, we emphasize that the estimates made in this section should be interpreted cautiously as they rely on several assumptions and do not take into account general relativistic effects of photon propagation.  
They are only meant to give an approximate range of luminosities and photon frequencies and need to be followed up with more detailed calculations.

\end{document}